\documentclass[pdftex]{article} 

\usepackage{tabularx}
\usepackage{arxiv}
\usepackage[utf8]{inputenc} 
\usepackage[T1]{fontenc}    
\usepackage{hyperref}       
\usepackage{booktabs}       
\usepackage{url}            
\usepackage{graphicx}
\usepackage{amssymb}   
\usepackage{lmodern}


\title{Can hot water discharged from industrial processes enhance the likelihood of waterspouts?}

\author{
 Valerio Capecchi \\
  Laboratorio di Meteorologia e Modellistica Ambientale per lo sviluppo sostenibile (LaMMA)\\
  Via Madonna del piano 10, Sesto Fiorentino\\
  Firenze (Italy)\\
  \texttt{capecchi@lamma.toscana.it} \\
   \And
 Bernardo Gozzini \\
  Laboratorio di Meteorologia e Modellistica Ambientale per lo sviluppo sostenibile (LaMMA)\\
  Via Madonna del piano 10, Sesto Fiorentino\\
  Firenze (Italy)\\
  \texttt{gozzini@lamma.toscana.it} \\
    \And
 Mario Marcello Miglietta \\
  Istituto di Scienze dell’Atmosfera e del Clima, Consiglio Nazionale delle Ricerche (ISAC-CNR)\\
  Corso Stati Uniti 4\\
  Padova (Italy)\\
  \texttt{m.miglietta@isac.cnr.it} \\
}

\begin{document}
\maketitle

\begin{abstract}
Italy and the surrounding seas are recognised as one of the European hotspots for
tornadoes and waterspouts. In recent years, the town of Rosignano Solvay (on the Northern Tyrrhenian coast) experienced repeated waterspouts affecting the same areas, raising local concern about the
possible influence of heated wastewater discharged into the sea by a nearby industrial site.
We reconstruct the mesoscale meteorological conditions of four intense waterspouts near Rosignano Solvay using a limited-area weather model at high-to-very-high resolution (inner domain grid spacing 500 m; sensitivity tests at 100 m). At the reported event times, the intensity of key mesoscale precursors (low-level wind shear, 1 km storm-relative helicity, maximum updraft intensity, and lifting condensation level) is consistent with the values typically associated with EF1 (or stronger) tornadoes and waterspouts. The model systematically predicts the peak of instability indices 2-3 hours earlier than the reported event times.
For one case study, we conduct two sea surface temperature sensitivity experiments to assess the potential atmospheric impact of heated wastewater discharge (temperature increases of +1.5 K and +5 K over a 10 km$^2$ area). The resulting changes in instability indices are marginal, with differences of at most 3\% relative to the control run.
%
A simple mass-balance estimate for the modified sea patch suggests that, given the reported discharge rates, a plausible impact of the warm water released from the industrial site could lead to an increase in the local sea surface temperature of approximately +0.7 °C over two months.
We conclude that synoptic and mesoscale conditions primarily govern waterspout initiation in this
region, while the direct effect of the small, warm coastal plume from the industrial discharge appears to be minor.
\end{abstract}


\section{Introduction}\label{sec:intro}
In recent years, Italy and the surrounding seas have been recognised as major European hotspots for tornadoes and waterspouts, as evidenced by numerous studies \cite{giaiotti2007climatology,antonescu2017tornadoes,miglietta2018updated,ingrosso2020statistical}. \cite{antonescu2017tornadoes} argue that tornadoes are an ``underestimated threat'' in Europe, because they are likely under-reported, due to a lack of a comprehensive pan-European tornado database, and because serious actions are not taken into consideration to prevent from their potential damage. Studies in Europe, including Italy, have examined synoptic patterns associated with tornadoes, often focusing also on mesoscale features \cite{romero2007european,matsangouras2014numerical,matsangouras2016study,taszarek2018climatological,ingrosso2020statistical,taszarek2020severe,bagaglini2021synoptic}. Based on the data presented in \cite{miglietta2018updated}, the Italian Tyrrhenian Coast stands out as particularly susceptible to such phenomena. In fact, between 2007 and 2016, this region witnessed an average annual density of approximately 2-3 tornadoes per 100 kilometres of its coastline. These relatively frequent events generally develop as waterspouts (ranging from 12 to 25 annually over the nearby sea) that later evolve  into tornadoes overland.

Over the ten-year long period 2007-2016, tornadoes across the broader Italian territory resulted in six fatalities, more than one hundred injuries, and estimated economic losses amounting to 80 million euros, as reported by \cite{miglietta2018updated} based on data from the European Severe Weather Database \cite{dotzek2009overview}. Such considerations emphasise the need for thorough investigations into their origins and dynamics. 

Research on tornadoes in Italy relies on both observations (recently provided in the framework of the storm report project; \url{www.meteonetwork.it/tt/stormreport}) and model analyses, providing insight into the formation and life cycle of tornado-spawning supercells \cite{miglietta2017effect,miglietta2017numerical,miglietta2020observational,avolio2021multiple,avolio2022tornadoes,de2025significant}.  Using ERA5 \cite{hersbach2020era5} global gridded reanalyses, \cite{bagaglini2021synoptic} focused on both the synoptic and mesoscale precursors that favour the formation and intensification of tornadoes in the Italian area. In line with previous scientific investigations \cite{rotunno1981evolution,weisman1986characteristics,matsangouras2016study,taszarek2017sounding,taszarek2020severe}, they found that positive anomalies of low-level vertical wind shear (LLS), convective available potential energy (CAPE) and storm relative helicity (SRH) in the lowest levels (below 3 km) occur over or nearby the regions where tornadoes or waterspouts are observed. In their study on the climatological properties of convective-prone environments, \cite{taszarek2020severe} showed that, over Europe, the mixed-layer CAPE is well correlated with the intensity of tornadoes, i.e., the higher the mixed-layer CAPE, the more severe the tornado. They also found that the low-level wind shear and SRH values are positively correlated with the severity of  tornadoes. As regards the sea surface temperature (SST) positive anomalies, some authors \cite{miglietta2017effect,avolio2021multiple,marin2021tornadoes} reported the crucial impact they have on tornado development, especially for tornadoes in the Ionian regions and, to a minor extent, in the Northern Adriatic area. \cite{miglietta2017effect} observed that even minor fluctuations in SSTs, such as an increase of 1 K across the whole Ionian Sea, may exert a substantial influence on the lifecycle of tornado-producing supercells. In particular, the authors found that the updraft helicity between 2 and 5 km in the presence of SST positive anomalies is approximately four times greater than that in the control simulation. Similarly, \cite{avolio2021multiple} conducted a sensitivity analysis employing a limited area weather model, focusing on four tornadoes in southern Italy. They increased uniformly the SST of the test experiment by 1 K with respect to the control simulation, where SSTs were taken from global data. Their findings indicated that positive SST anomalies contribute to a warmer and more humid atmosphere in the lower atmospheric layers, resulting in heightened CAPE and consequently increasing the potential for stronger updrafts. Some studies conducted outside the Mediterranean context have also investigated the impact of sea surface temperature on the development and intensification of extreme convective events. For example, \cite{sari2023understanding}, in their analysis of a severe hailstorm event in Indonesia, performed sensitivity experiments by increasing SST by 3 K in the vicinity of the area of interest. They found that, relative to the control simulation, the perturbed experiment led to enhanced surface turbulent fluxes (both latent and sensible heat), thereby increasing low-level moist static energy and CAPE. This thermodynamic enhancement promoted stronger updrafts and contributed to a more intense convective response.

Several studies highlight the significant role of orography in tornado outbreaks. To stick to the European context, \cite{mateo2009study} identified that orography can modify convergence lines, making them the primary trigger for a tornado that struck northeastern Spain in 2006. Similarly, \cite{rigo2022observational}, \cite{matsangouras2014numerical} and \cite{homar2003tornadoes} studied the interactions between topographic features and the onset and life-cycle of tornado-producing storms. In their analysis of the complex orographic setting of the Pyrenees, \cite{rigo2022observational} suggest that abrupt changes in terrain elevation along the trajectory of a tornado-producing cell may contribute substantially to the initiation of tornadogenesis. Specifically, they document instances where the cell experienced an elevation gain of approximately 200 m within 12 minutes of its movement. The authors of \cite{matsangouras2014numerical} performed sensitivity experiments by varying the topographic height in their numerical simulations; they found that decreasing terrain elevation resulted in reduced instability variables relevant to tornadogenesis, in some cases by as much as 100\%. The study by \cite{homar2003tornadoes} demonstrated that the interaction between orographically forced flows and the timing of solar radiation played a crucial role in establishing a favourable environment for tornadogenesis during a severe weather event over complex terrain. Additionally, \cite{miglietta2017numerical} conducted high-resolution numerical simulations to reconstruct the mesoscale meteorological conditions that led to a multivortex EF3 tornado striking southern Italy in 2012. By performing two sensitivity tests in which the orography was reduced by 50\% and 80\%, the authors found that the tornado-spawning supercell became weaker or was strongly suppressed, respectively. According to their study, orographic features initiated convection cells that subsequently moved over the sea, and were maintained by moisture and heat in phase with boundary layer rolls.

Previous research demonstrates that the extremely fine resolution required to simulate tornadoes numerically makes it challenging to reproduce their evolution in real-time. Typically, numerical simulations of tornadoes are conducted retrospectively under highly idealised conditions and at very high-resolution for idealised runs \cite{rotunno2016axisymmetric}.
In real-time applications, limited-area models may be employed to identify the conditions favourable for their development using an ensemble approach, such as in the NCAR real time ensemble forecast (\url{https://ensemble.ucar.edu}, last visit in December 2025).

\begin{figure}
    \includegraphics[width=14cm]{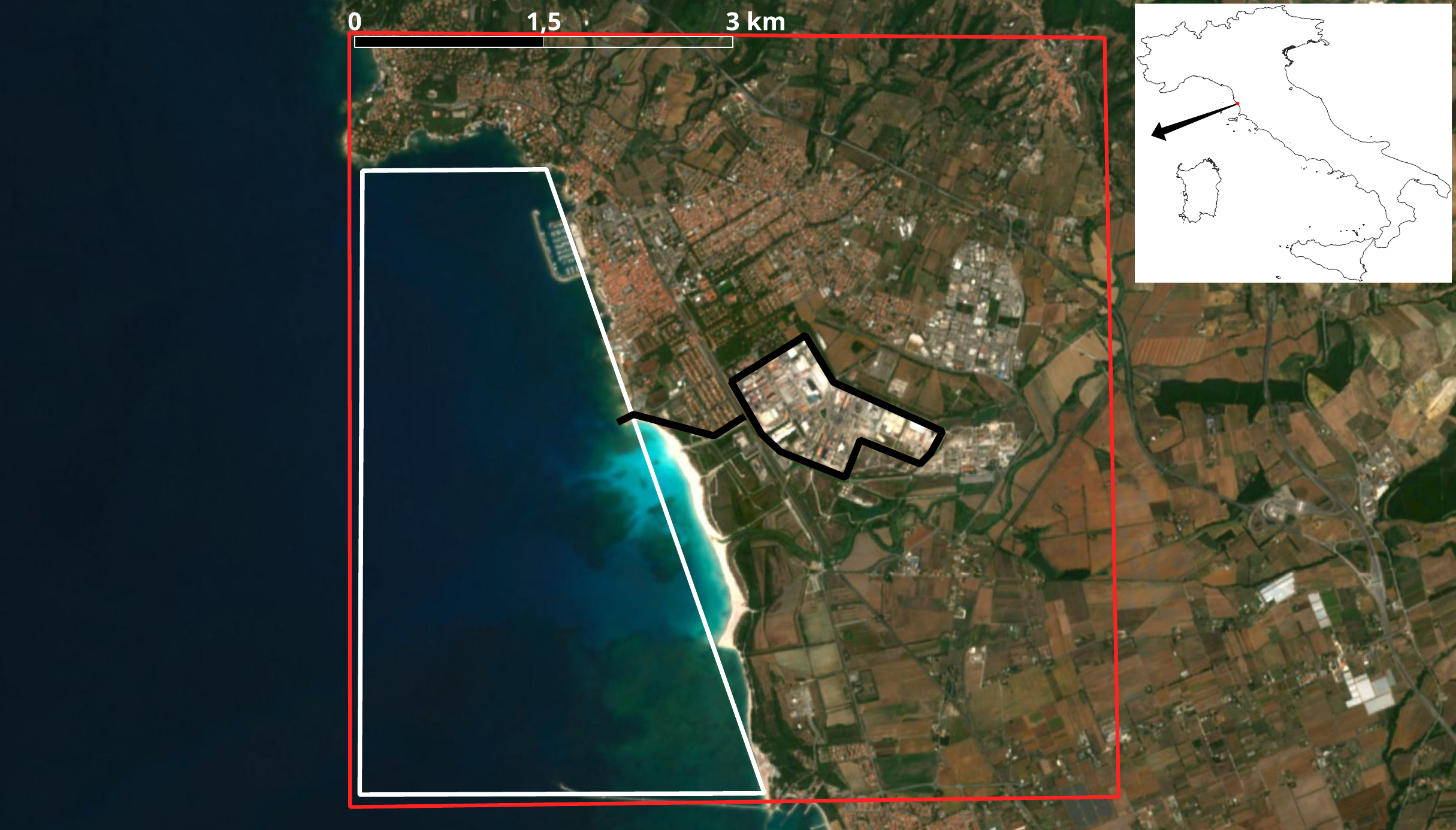}
    \caption{Zoom on the Rosignano Solvay area belonging to the Rosignano Marittimo municipality. The red square delineates the integration domain for the higher-resolution Meso-NH simulation, which employs a grid spacing of 100 m and is applied to the 25SEP2020     case only. The white trapezoid indicates the region where sea surface temperatures were modified in the sensitivity experiments. Such trapezoid  covers an approximate area of 11 km$^2$. The black line represents the path of the ditch that discharges water from the industrial facility, which is indicated with the black polygon and is located less than 1 km inland from the coastline. Satellite image from Sentinel-2 cloudless \textcopyright\ EOX IT Services GmbH.}\label{fig:domain100}
\end{figure}

As regards the areas most affected by tornadoes, the town of Rosignano Solvay in the Northern Tyrrhenian Coast of Italy (belonging to the Rosignano Marittimo municipality, see the inset map in Figure \ref{fig:domain100}) stands out.
In less than a decade, the town has experienced four waterspouts, two of them directly impacting the same buildings. The growing concern prompted the municipality to seek insights from the local weather agency (i.e., LaMMA, Laboratorio di Meteorologia e Modellistica Ambientale per lo sviluppo sostenibile) regarding the potential influence of local factors that could have contributed to a higher frequency of waterspouts. In particular, the request was to assess the impact of warm waters along its coastline on the occurrence of waterspouts. Notably, the Solvay industrial site, located just a few hundred meters inland from the Rosignano coast (see the black polygon in Figure \ref{fig:domain100}), releases heated water, utilised in its production processes, through a small channel connected to the sea (see the black segment in Figure \ref{fig:domain100}). Based on data supplied by the Solvay administration, the mean wastewater temperature in 2020 was 26.2$^\circ$ C (approximately 7$^\circ$ C warmer than the observed annual mean of SST), and the drainage ditch released an average of 8.7 $m^3 h^{-1}$ of water into the sea.

\begin{figure}
        \includegraphics[width=14cm]{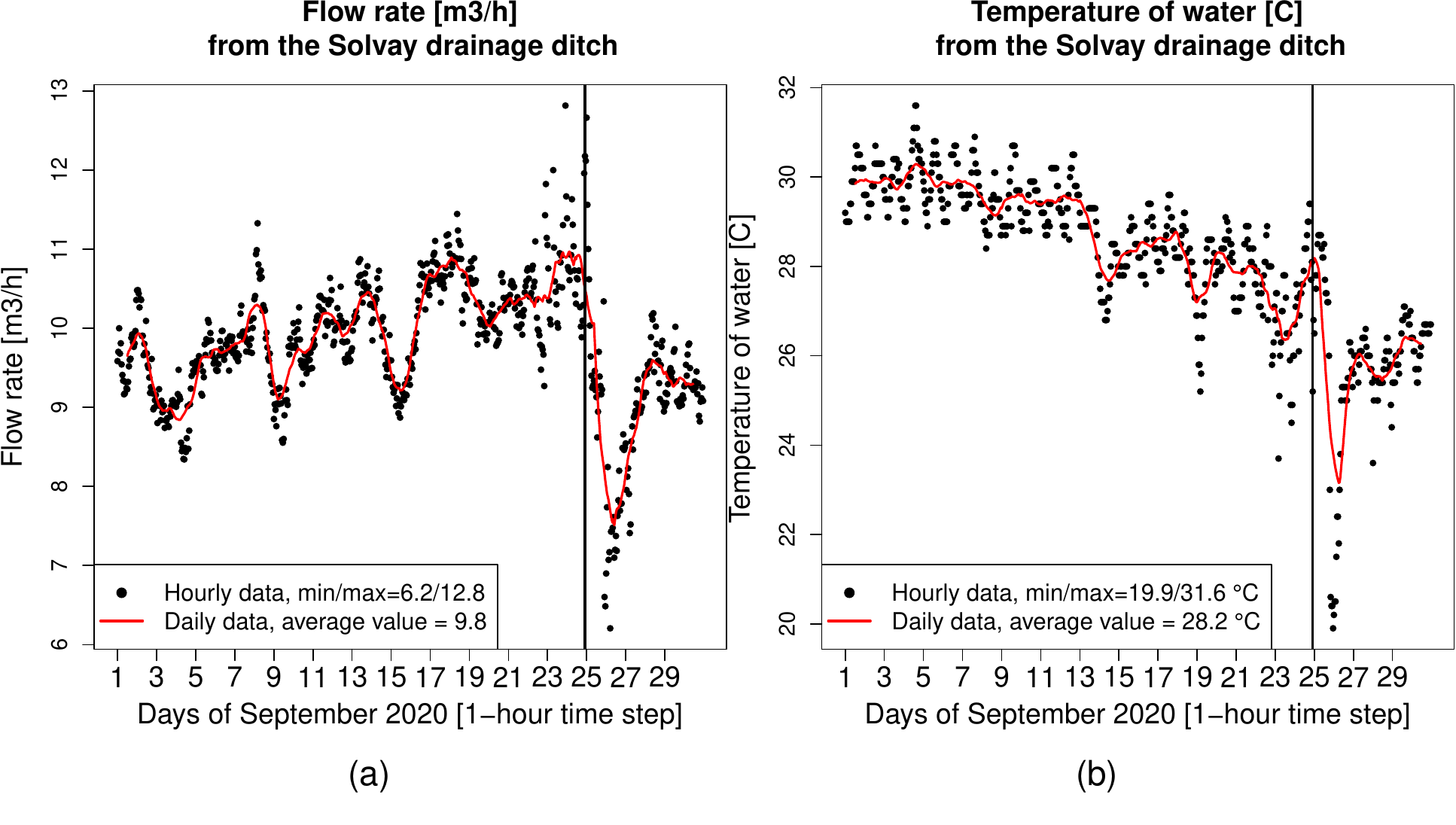}
     \caption{1-30 September 2020 hourly data from the Solvay Chimica Italia discharge ditch: (a) flow  (unit of measure is ${m^3}$/{hour}) indicated with black triangles and (b) water temperature (unit of measure is degree Celsius) indicated with black circles. In both plots the red line indicates the 24-hour moving average and the black vertical line indicates midnight of the 25th of September 2020. Data courtesy of Solvay Chimica Italia.}\label{fig:Solvay}
\end{figure}

One of the main objectives of our study is to simulate numerically the conditions conducive to the four waterspout events and evaluate whether the model outputs align with the atmospheric conditions typical of extreme phenomena, such as waterspouts. This goal is achieved by reconstructing the meteorological conditions conducive for tornadoes/waterspouts with a numerical weather model run at high-resolution. A further objective is to assess whether the discharge of residual warm water from industrial processes into the sea, leading to locally increased SSTs, may enhance the likelihood of waterspout formation and development. This analysis is conducted for one of the four cases only, due to the availability of detailed information on the flow rate and temperature of the discharged warm water provided by the management of the industrial site (see data shown in Figure \ref{fig:Solvay}).
We employed numerical simulations using the Meso-NH model \cite{lac2018overview}, incorporating both actual SSTs and modified SSTs that estimate the influence of wastewater released into the sea near Rosignano Solvay. Our research also aims to address the question of whether a limited body of water (of the order of a few square kilometres, i.e., less than 15 km$^2$) with a relatively warmer surface temperature than the surrounding area can trigger or increase the frequency of waterspous in the surrounding region. To our knowledge, only \cite{avolio2022tornadoes,avolio2023comparative} have conducted in-depth studies using high-resolution numerical models to simulate conditions favourable to tornado development in the Tyrrhenian coast. Their research revealed that these events frequently occur at the boundaries of convective systems in areas of low-level convergence, also providing indications for the conditions that favour the formation of tornadoes in the region. 

The paper is organised as follows: Section \ref{sec:data} presents a concise overview of the four study cases, highlighting their shared and distinctive characteristics from a synoptic point of view. Additionally, it details the experimental design of the Meso-NH model used for numerical reconstruction  of the environmental conditions conducive to their occurrence. The results are presented and analysed in Section \ref{sec:results} and further discussed in Section \ref{sec:discussion}. In the last Section, we discuss the limitations of the approach used and provide a possible explanation of the mechanisms and factors that determine the origin and development of tornadoes in the Rosignano Solvay area.

\section{Data and methods}\label{sec:data}
\subsection{Description of the waterspout cases}\label{ssec:desc}
In November 2020, the municipality of Rosignano Marittimo (located on the Northern Tyrrhenian Coast, see Figures \ref{fig:domain100} and \ref{fig:italy})
asked LaMMA to shed light on four waterspouts that were observed in its territory. These extreme weather events occurred on the following dates, with approximate times indicated:
\begin{itemize}
\item 17 December 2011 early morning, hereinafter the 17DEC2011  case;
\item 27 November 2012 11:45 UTC, hereinafter the 27NOV2012   case;
\item 10 September 2017 2:30 UTC, hereinafter the 10SEP2017    case;
\item 25 September 2020 18:45 UTC, hereinafter the 25SEP2020     case.
\end{itemize}

Before delving into the specific characteristics of each case, some common elements can be identified across the four events. These include the persistence of the jet stream, pronounced baroclinicity, and strong vertical wind shear. In all cases, a cold front was present, either propagating south-eastward or quasi-stationary, but consistently located in close proximity to the event. The pre-frontal environment was characterized by high humidity, marked low-level wind convergence (between westerly and northwesterly flows or between southerly and southwesterly flows), and elevated CAPE values. To visualise some of such common features, maps from ERA5 data of 250-hPa jet, geopotential height at 500-hPa with superimposed temperature, deep level shear in the 0-6 km layer are provided in Appendix \ref{app:fig1} in Figures \ref{fig:250jetERA5}-\ref{fig:dlsERA5}, respectively.
%
\begin{figure}[t]
\includegraphics[width=14cm]{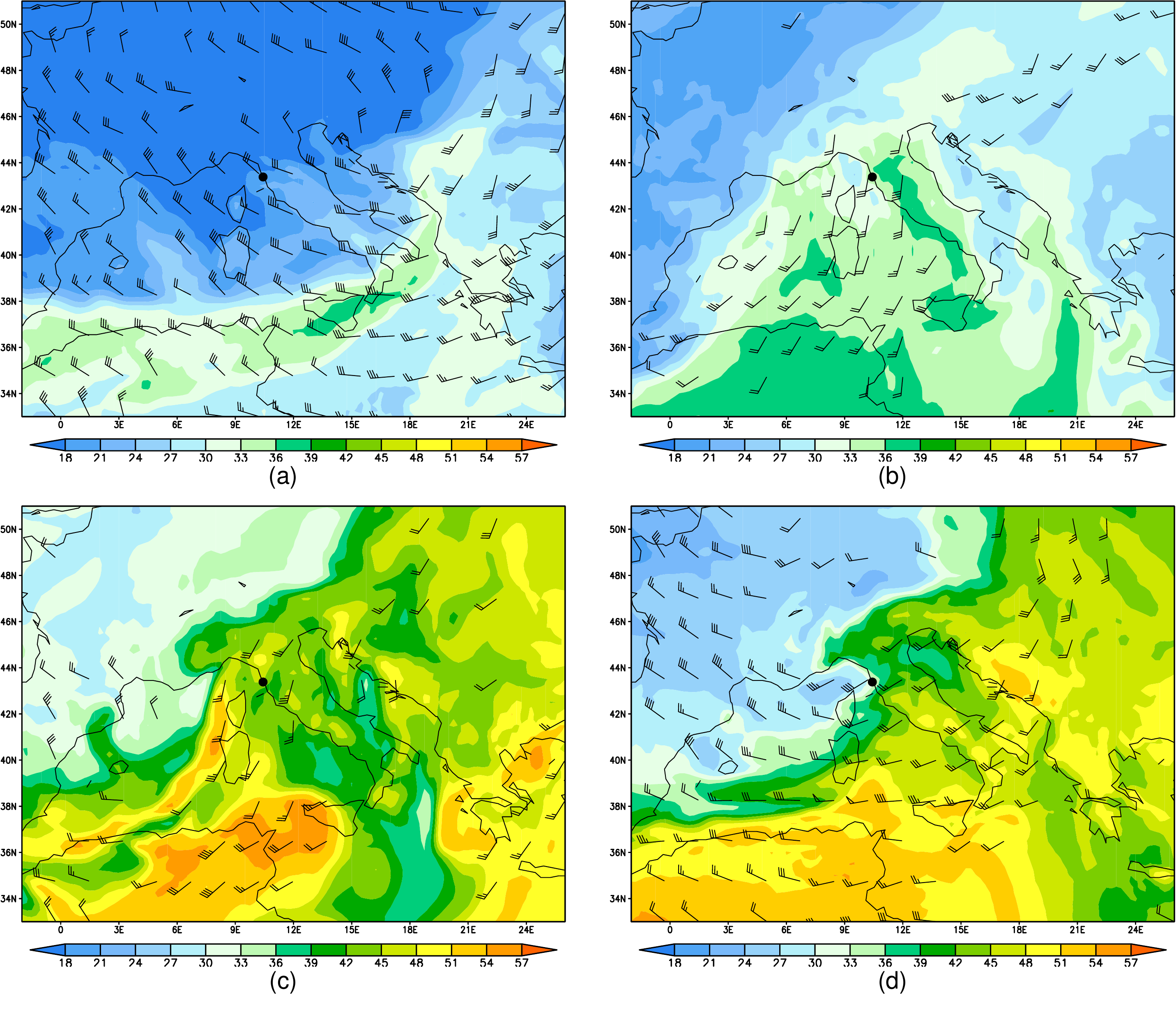}
\caption{850-hPa equivalent potential temperature, $\theta_E$ (unit of measure degree Celsius), and wind speed (unit of measure knots) and direction from ERA5 data at the approximate time of the waterspout for the 17DEC2011  (a), 27NOV2012   (b), 10SEP2017    (c) and 25SEP2020     (d) case.  Barbs are shown for wind speed greater than 20 knots. The approximate location of Rosignano Solvay is indicated with the black point ($\bullet$).}\label{fig:thetaE}
\end{figure}

Two cases (17DEC2011  and 25SEP2020) are characterised by waterspouts (later making landfall) formed in conjunction with thunderstorms developing during or after the passage of a cold front. To identify these fronts, Figures \ref{fig:thetaE}a and \ref{fig:thetaE}d show a distinct boundary in the 850-hPa equivalent potential temperature ($\theta_E$) field, where colder and drier air (lower $\theta_E$ values) replaces warmer and moister air (higher $\theta_E$ values). This feature is particularly pronounced in the 25SEP2020 case. Specifically, Figure \ref{fig:thetaE}d reveals in the area of interest a narrow, sloping zone characterized by strong $\theta_E$ gradients, with values decreasing from 39-42$^\circ$ C to 27-30$^\circ$ C over a distance of less than 200 km. Further, a convergence line in the lower atmospheric layers formed between westerly and northwesterly winds, which favoured the initiation of thunderstorms (maps not shown). The remaining two cases (27NOV2012 and 10SEP2017) are characterised by the presence of an intense and very humid meridional flow (Figures \ref{fig:thetaE}b and \ref{fig:thetaE}c). Unlike the first two cases, thunderstorms developed in the prefrontal phase, i.e., before the actual cold front entered, as the affected area was still in the warm sector ($\theta_E$ values in the range 33-39$^\circ$ C for 27NOV2012   and 39-45$^\circ$ C for 10SEP2017, see Figures \ref{fig:thetaE}b and \ref{fig:thetaE}c, respectively). However, in both cases, the $\theta_E$ patterns suggest the intrusion of relatively colder air (in terms of $\theta_E$) at more local scale, which may have displaced the pre-existing warm air and initiated convection, subsequently leading to the development of thunderstorms. Maps of 10-metre wind speed and direction (not shown) indicate a less pronounced, yet still discernable, convergence lines near the surface formed between southerly and southwesterly winds, that might have contributed to the initiation of
convection.

To characterise better the four cases, we provide some observational evidences, on the basis of data available in the European Severe Weather Database (ESWD, \url{https://eswd.eu}, last access 4 September 2025), managed by the European Severe Storm Laboratory \cite{dotzek2009overview}. In assessing the damages caused by the extreme events, we followed the classification adopted by ESWD, which is based on the International Fujita (IF) Scale for tornado and wind damage assessment \cite{Groenemeijer2023international}.

%
\subsubsection{The 17DEC2011 \ case}\label{sssec:undici}
\begin{figure}[]
\includegraphics[width=14cm]{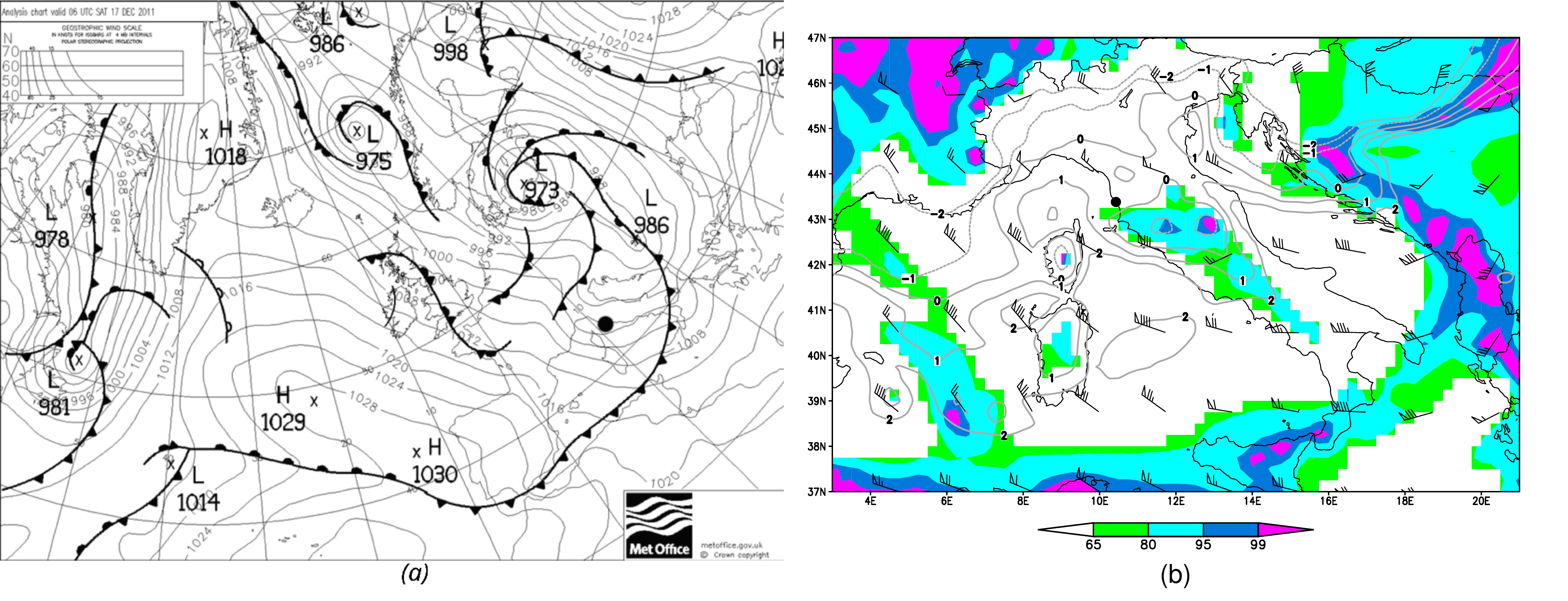}
\caption{17DEC2011  case: (a) surface pressure reduced at sea level and fronts at 06 UTC (plot courtesy of UK MetOffice) and (b) 850-hPa wind speed and direction (barbs), relative humidity (shaded colours) and temperature (white contour) at 06 UTC  (ERA5 data). The approximate location of Rosignano Solvay is indicated with the black point ($\bullet$).}
\label{fig:casoundici}
\end{figure}

The ESWD has documented a tornado event in the Rosignano area categorised as IF0.5 intensity scale, indicating an instantaneous wind speed of up to 33 $m/s$. This report holds a quality control designation of QC1, signifying its confirmation by a reliable source within a voluntary observer network or a national hydro-meteorological service, as per ESWD guidelines. Roof and ship damages have been reported in the Rosignano Solvay area, along with several large tree branches broken. Furthermore, a bus was overturned off the road in the vicinity. For this case, a red alert code for coastal storm surge was issued by the local Civil Protection. The red code is the highest level foreseen by Civil Protection procedures and has been issued only eight times  in the period 2011-2023. Figure \ref{fig:casoundici}a illustrates the surface pressure map at 06 UTC 17 December 2011, that shows a pronounced gradient between the low pressure centre (973 hPa) located over North Germany and Poland and a wide high pressure area (1030 hPa) offshore the Portugal coast. Figure \ref{fig:casoundici}b illustrates the wind, temperature and relative humidity values at 850-hPa pressure level from ERA5 dataset at 06 UTC 17 December 2011. The data indicate that at the time of the waterspout occurrence, the low levels of the atmosphere (at least up to 850-hPa pressure level) were characterised by strong northwesterly winds (wind speed up to 70 $km/h$), pushing towards the Tyrrhenian Coast dry (relative humidity approximately 40\% or less) and cold air (temperature approximately 1$^\circ$ C). Such conditions suggest that the area of interest is likely affected by air following the passage of a cold front.

%
\subsubsection{The 27NOV2012  \ case}\label{sssec:dodic}
\begin{figure}[]
\includegraphics[width=14cm]{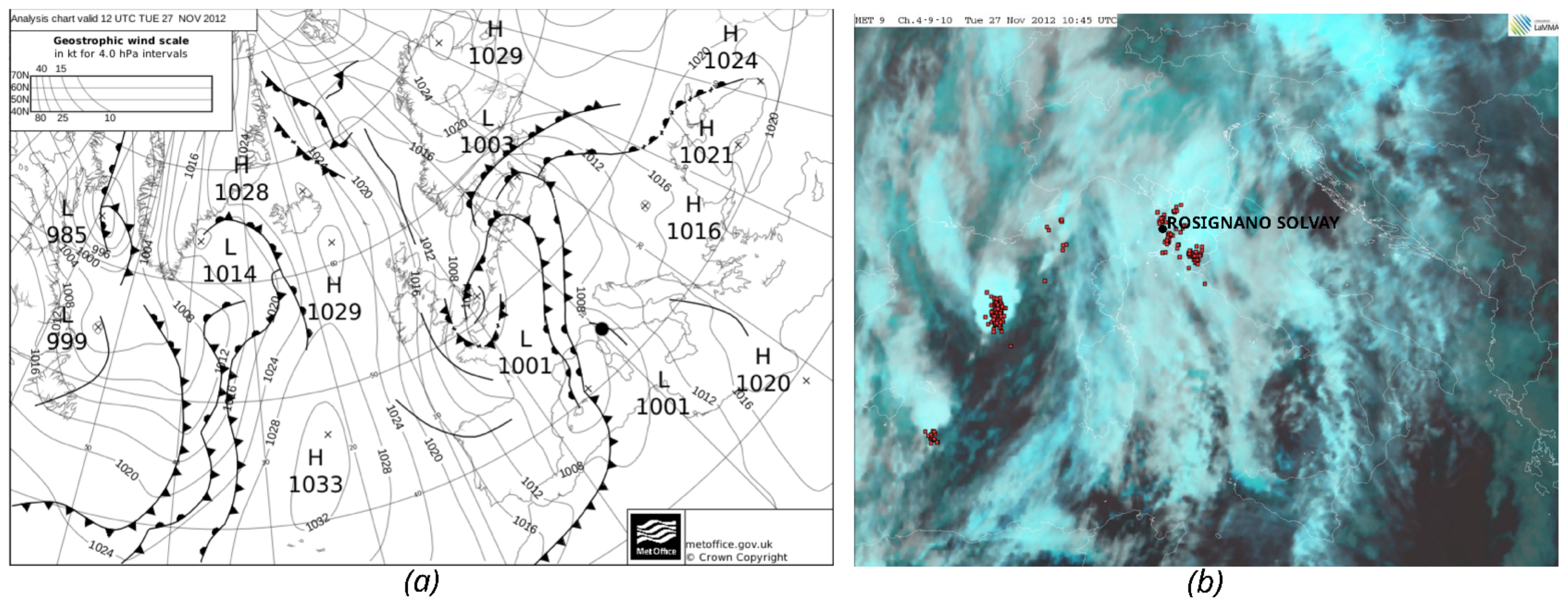}
\caption{27NOV2012   case: (a) surface pressure reduced at sea level and fronts at 12 UTC (plot courtesy of UK MetOffice) and (b) Meteosat infrared image at 10.45 UTC with superimposed lighting strikes observed by the Blitzortung network. The approximate location of Rosignano Solvay is indicated with the black point ($\bullet$).}
\label{fig:casododici}
\end{figure}

The ESWD database reports the occurrence of a tornado event in the municipality of Rosignano Marittimo. Based on information from a website, photos, videos, and a newspaper report, the event has been classified as IF1 intensity scale. Such events are associated with instantaneous wind speed equal to $40\pm8$ $m/s $. The event was associated with heavy rain and a funnel cloud was observed; the quality control assigned to this report is QC1. At 12 UTC, the surface pressure map (Figure \ref{fig:casododici}a) reveals the presence of a low pressure centre (1001 hPa) located close to the Sardinia Channel, which activates southerly flows over the Tyrrhenian Sea. A troughs is clearly visible in the Rosignano Solvay area, possibly associated with unstable warm air beneath cold air. The deep-level wind shear, computed between the 500- and 1000-hPa levels, is 18-20 $m/s$ and the low-level wind shear computed between the 900- and 1000-hPa levels, is 10-12 $m/s$ (ERA5 data, not shown); in both cases limited time changes in wind direction are evident. High deep- and low-level wind shear were noted to be favourable to tornado development by \cite{colquhoun1989objective} and \cite{thompson2003close} in US and by \cite{bagaglini2021synoptic} and \cite{ingrosso2020statistical} in Italy. Figure \ref{fig:casododici}b shows the Meteosat Second Generation infrared image (channel 10 corresponding to a wavelength of 12 $\mu m$), with superimposed lightning strikes observed by the Blitzortung network (\url{www.Blitzortung.org}, last accessed on 3 September 2025). The image indicates a strong convective activity near the coast of Tuscany at the time when the waterspout was observed.

%
\subsubsection{The 10SEP2017   case}\label{sssec:dicia}
\begin{figure}[]
\includegraphics[width=14cm]{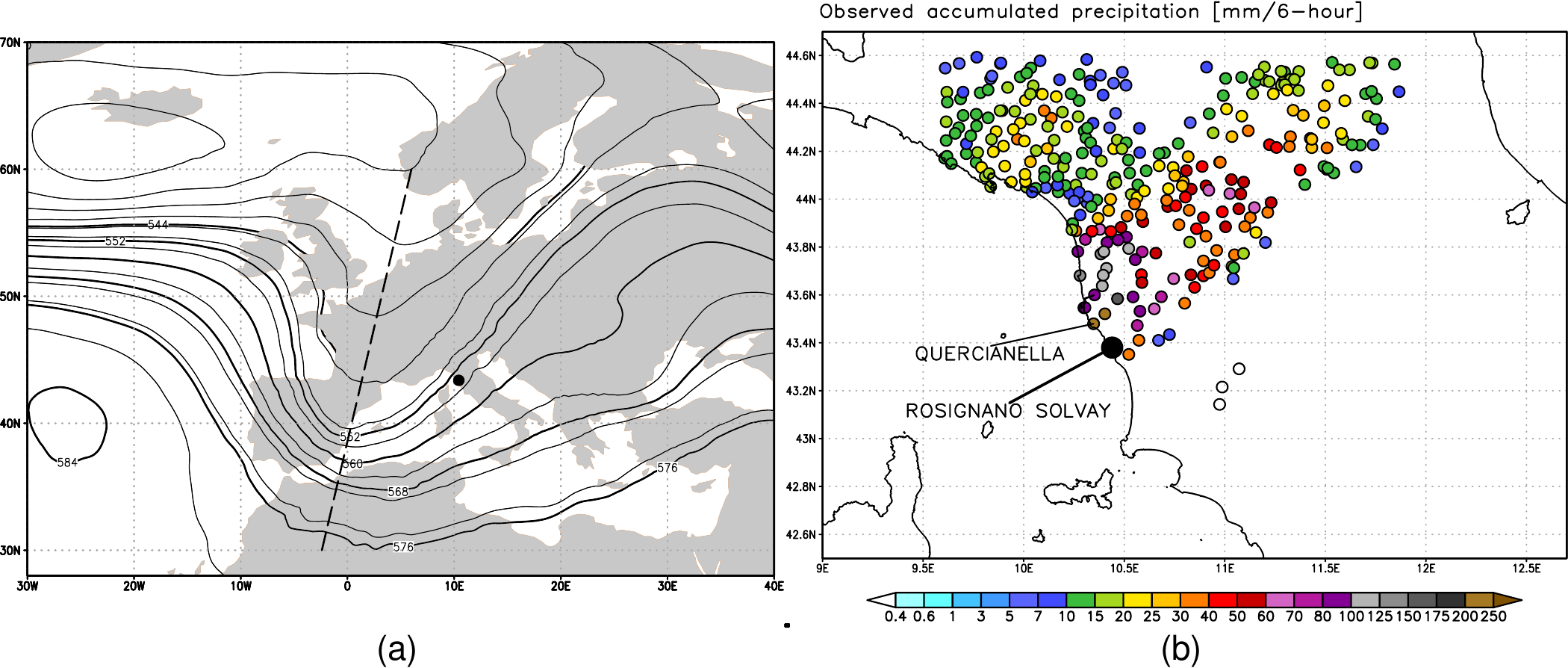}
\caption{10SEP2017    case: (a) geopotential height at 500 hPa in decametres at 00 UTC (ERA5 data) and (b) observed rainfall accumulated in the 6-hour period ending on 10 September 2017 at 03 UTC, registered by automatic weather stations in the Tuscany region and surrounding areas. The approximate location of Rosignano Solvay is indicated with the black point ($\bullet$).}
\label{fig:casodicia}
\end{figure}

The 10SEP2017 case is well known and studied in the scientific literature \cite{federico2019impact,lagasio2019effect,capecchi2021assimilating}. The ESWD database reports no records for this event in the Rosignano area; however, heavy rain associated with a strong low-level wind convergence line was observed in the broader area of interest \cite{federico2019impact,capecchi2021assimilating}. The automatic weather station of Quercianella, which is located approximately 10 km north of Rosignano Solvay (Figure \ref{fig:casodicia}b), registered more than 200 mm in less than 6 hours with an estimated return period of more than 200 years for 3-hour duration rainfall events \cite{capecchi2021assimilating}. Intense precipitation was accompanied by significant lightning activity \cite{federico2019impact}. Such event was associated with the slow movement of a trough, whose axis was oriented from the Scandinavian Peninsula to the Western Mediterranean Sea (subjectively identified with the dashed line in Figure \ref{fig:casodicia}a). This determined the deepening of a low pressure area, in the lee of the Italian Alps. A well defined line of convergence of low-level winds between southerlies and southwesterlies over the Tyrrhenian Sea acted as a trigger to overcome the moderate convective inhibition  \cite{capecchi2021assimilating} and to release the instability (CIN values less than 100 $J\cdot kg^{-1}$ and CAPE values in the range 500-1000 $J\cdot kg^{-1}$, ERA5 data not shown here). For additional and mode detailed descriptions of the synoptic conditions contributing to this notable event, readers may refer to the cited studies \cite{federico2019impact,capecchi2021assimilating}.

%
\subsubsection{The 25SEP2020     case}\label{sssec:venti}
\begin{figure}[]
\includegraphics[width=14cm]{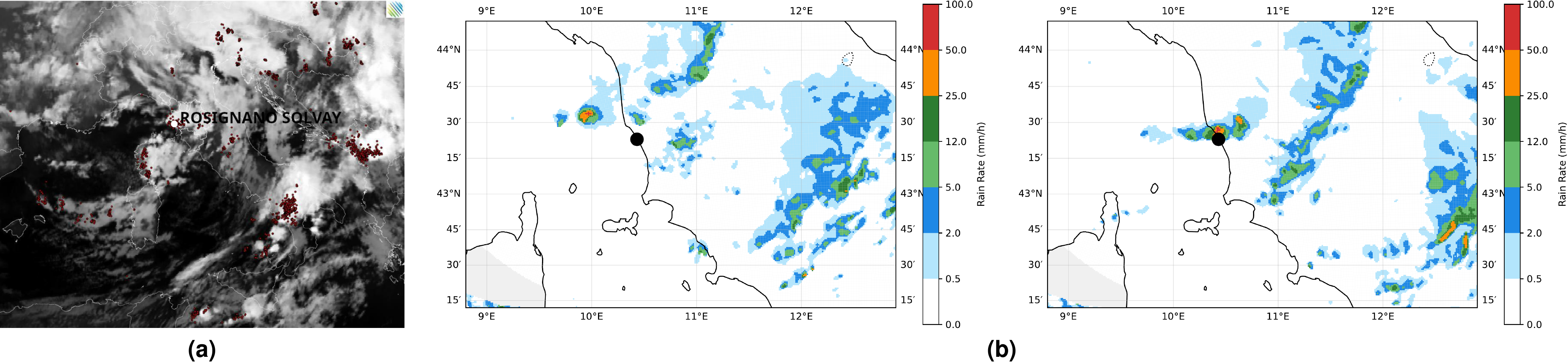}
\caption{25SEP2020     case: (a) Meteosat infrared image at 18.45 UTC with superimposed lighting strikes indicated with the filled red squares and (b) surface rainfall intensity estimated by radar data at 18.30 UTC (left) and 19:00 UTC (right). Data retrieved from the archive described in \cite{franch2026itdpc}, released under the CC BY-SA 4.0 license. The approximate location of Rosignano Solvay is indicated with a white point in panel (a) and with a black point in panel (b).}
\label{fig:casoventi}
\end{figure}

For this date, the ESWD dataset contains three records in the area of interest. The first reports a ``tornado/waterspout'' event in the Rosignano Marittimo area at 18:45 UTC on September 25,  2020. It is based on information from photos and videos, an eye-witness report of the damage and a trained storm spotter observation. Its intensity is classified as IF2, with eight people injured. The instantaneous wind intensity associated with such events is in the range $60\pm18$ $m/s$. The second record reports  ``large hail'' in the same location 30 minutes later. The third record, that locates the event near the town of Vada (less than 3 km south of the town of Rosignano Solvay), reports a ``tornado/waterspout'' at 19:15 UTC, which caused several damages at ships anchored in the local harbour. Its intensity is classified as IF1 with a quality control code QC1. For each of the three phenomena, there are links to videos and photographic documents available on ESWD. Moreover, numerous  observations corroborate the extent and severity of the damage. Satellite and radar images depict the formation of an intense precipitation core over the sea near Livorno, located 20 km to the north of Rosignano Solvay, around 18:45 UTC (see Figures \ref{fig:casoventi}a and \ref{fig:casoventi}b). Subsequently, this core moves southeastward, impacting the Rosignano area around 19:00 UTC. Radar data shown in Figure \ref{fig:casoventi}b suggest that the phenomena tend to align along a convergence line formed by southwesterly and northwesterly winds. Additionally, the red core visible in Figure \ref{fig:casoventi}b, and corresponding to an estimated rainfall intensity rate of approximately 50 $mm/h$, indicates very strong convection, possibly associated with intense phenomena, such as tornadoes. The observed  movement of the intense cell further indicates that the tornado in the Rosignano Solvay area initially formed as a waterspout, extending into the coastline for a few hundred meters before dissipating.

%
\subsection{Experimental design of the Meso-NH model for the numerical reconstruction of the four  events}
Numerical reconstructions of the four severe weather events were realised by using the Meso-NH model, version 5.4.3, which was released in March 2020. Developed by CNRM (Meteo-France/CNRS) and the Laboratoire d'A\'erologie at the University of Toulouse, this model aims to explore atmospheric phenomena across different scales: from the scale typical of planetary waves (horizontal length approximately 2000 km or more) to meso-scale (2000-2 km) down to micro-scale (2000-2 m). For an in-depth understanding of the model and its wide-ranging applications, refer to \cite{lac2018overview}. Initial and boundary conditions were derived from global operational analyses and forecasts provided by the European Centre for Medium-Range Weather Forecasts (ECMWF), whose horizontal grid spacing is set to 0.125$^\circ$ for all the four events. We acknowledge that this setting may affect the quality and reliability of the high-resolution simulations, as the ECMWF global model (Integrated Forecasting System, IFS) has undergone several changes over the years. These changes affect not only the model's physical parameterisations but also its spectral resolution. For instance, the horizontal resolution of the IFS used for operations in 2017 and 2020 was approximately 9 km, whereas it was approximately 16 km in 2011 and 2012. However, we interpolated the global data to an intermediate resolution to ensure the simulation forcing remains comparable among the events. This configuration was also shown to be effective in previous studies that employed IFS data and the Meso-NH model \cite{capecchi2020reforecasting,capecchi2021reforecasting}. Short-term simulations (forecast length shorter than 24 hours) were conducted for each of the four events using the Meso-NH model in two nested grids with grid spacing of 2500 m and 500 m, respectively.
\begin{figure}
\includegraphics[width=14cm]{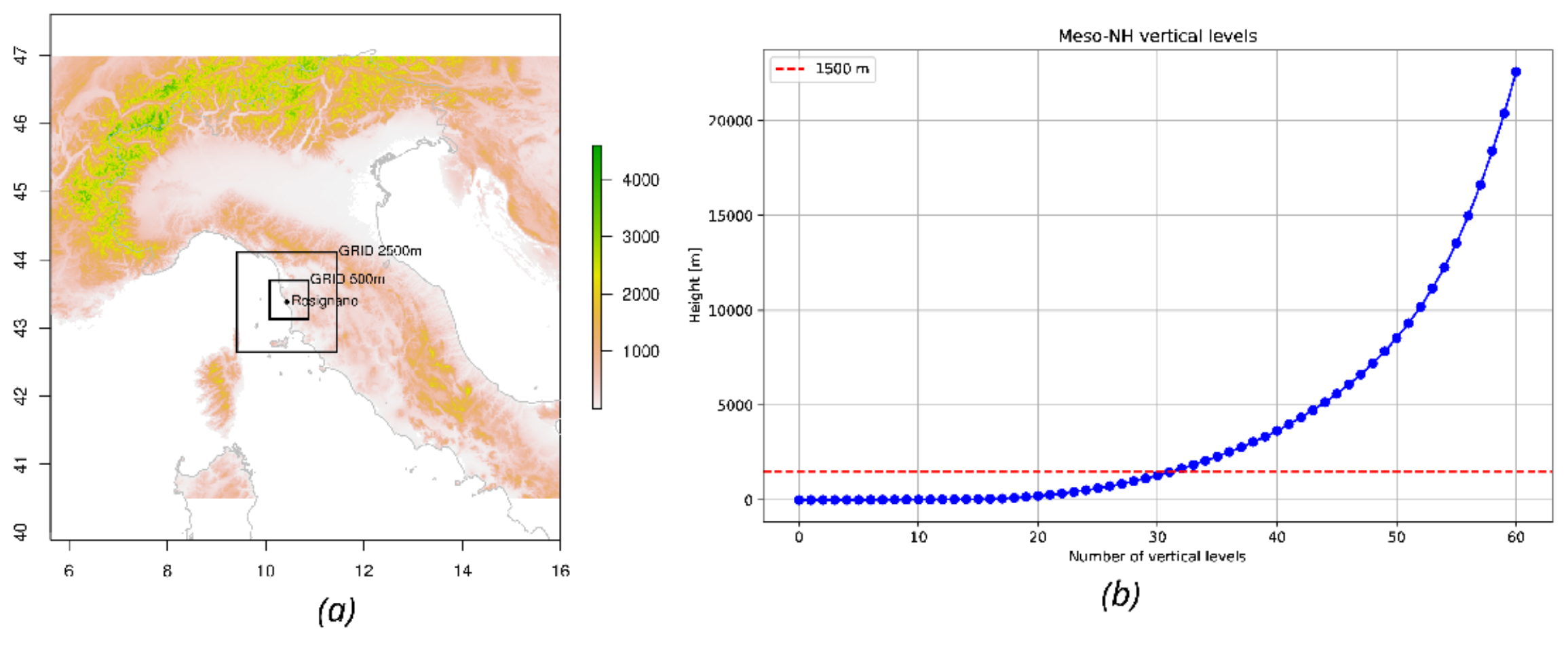}
	\caption{(a) Location of the town of Rosignano and the two Meso-NH domains overlaid on model topography. (b) Distribution of the vertical levels (blue line with dots) for both Meso-NH grids; the red dashed line indicates the 1500 m height.}\label{fig:italy}
\end{figure}
See Table \ref{tab:startdates} for actual start and end times of each simulation.
\begin{table}[]
  \caption{Dates and times of the four waterspouts events occurred in the area of Rosignano Marittimo municipality and start and end times of numerical reconstructions; hours are expressed in UTC. Approximate time of waterspout occurrence is based on the reports available from the ESWD database. Since for the 17DEC2011  case, no time is reported (the generic ``early morning'' time window is indicated), we, subjectively, picked 04:00.}\label{tab:startdates}
  \begin{tabularx}{\textwidth}{llll}
  \toprule
        \textbf{Date of} & \textbf{Approximate time} & \textbf{Start time} & \textbf{End time} \\
        \textbf{event} & \textbf{of waterspout} &  &  \\
        & \textbf{occurrence} &  &  \\
  \midrule
  2011-12-17 & 04:00 & 00:00 & 09:00\\
  2012-11-27 & 10:45 & 00:00 & 15:00\\
  2017-09-10 & 02:30 & 00:00  & 06:00\\
  2020-09-25 & 18:45 & 12:00  & 24:00\\
  \bottomrule
  \end{tabularx}
\end{table}
The location of the two grids in the wider Italian area are shown in Figure \ref{fig:italy}a. The Meso-NH outer domain has an extension of approximately 27000 km$^2$ and a grid spacing equal to 2500 m; it is squared with a number of points equal to 66 in each direction. The inner Meso-NH domain has an extension of approximately 4000 km$^2$ and a grid spacing equal to 500 m; it is squared with a number of points equal to 127 in each direction. Model outputs were saved every 15 minutes.  Key model settings are summarised in Table \ref{tab:mesonh}, aligning with those employed in \cite{capecchi2021reforecasting} and used for the reconstruction of two extreme precipitation cases occurred in Italy. Convective processes (both deep and shallow) are not parameterised in either grid. Vertical discretization involves 60 levels distributed from the surface to approximately 22 km, with 31 levels in the first 1500 meters (Figure \ref{fig:italy}b). As regards the advection scheme used for horizontal and vertical velocities, the centred discretization of fourth order centred on space and time (CEN4TH) scheme \cite{lac2018overview} is used. For the time integration, the Runge-Kutta fourth-order (RKC4) method is deployed; it was shown  \cite{ricard2013kinetic} that the coupled option CEN4TH and RKC4 allows an effective resolution \cite{skamarock2004evaluating} of the order 4-6$\times\Delta x$, where $\Delta x$ is the mesh size of the model. With such settings, to assure the numerical stability of the simulations, time steps equal to 6 s and 1.5 s are used for the outer and inner grid, respectively.
\begin{table}
\caption{Meso-NH model settings.}
\label{tab:mesonh}
  \centering
  \begin{tabular}{ll}
    \toprule
\textbf{Scheme} & \textbf{Reference} \\
    \midrule
Radiation      & ECMWF radiation scheme \cite{morcrette2008impact} \\
Boundary layer & Bougeault and Lacarrere \cite{bougeault1989parameterization} \\
Microphysics   & ICE3 scheme \cite{caniaux1994numerical} \\
Land-surface   & SURFEX model \cite{masson2013surfex} \\
Turbulence     & 1.5-order TKE equation \cite{cuxart2000turbulence} \\
    \bottomrule
  \end{tabular}
\end{table}

%
%
For the case study 25SEP2020     a higher-resolution domain (grid spacing equal to 100 m) is nested within the 500 m grid; it is square-shaped, with 62 points in each direction covering an area of almost 40 km$^2$ (see the red square in Figure \ref{fig:domain100}). Such additional inner grid shares the schemes listed in Table \ref{tab:mesonh} with its parent grid, except for the turbulence scheme. In this grid, no turbulence scheme is applied, as up to 90\% of the turbulence energy is resolved directly according to \cite{lac2018overview}. The time step is set to 0.5 s. To represent the albedo of the clear waters discharged from the drainage channel of the Solvay industrial plant, the model's albedo in the sea portion facing Rosignano (white trapezoid in Figure \ref{fig:domain100}) was set uniformly equal to 0.135.
In addition to the control simulation (hereafter referred to as \verb+CNTRL+), where initial conditions are interpolated from global model data, two SST sensitivity tests were conducted. In the first test (hereafter referred to as \texttt{SST15}), the SST in the sea area outlined by the white trapezoid shown in Figure \ref{fig:domain100} was increased by 1.5 K compared to the mean value of sea surface temperature reported by the large-scale model. The choice of a 1.5 K SST increment is inspired by the findings reported in \cite{miglietta2017effect}, where variations in SST values of $\pm$ 1 K have significant nonlinear effects on the vertical velocity of a tornado-spawning supercell. The decision to raise the sea surface temperature by 1.5 K in the sea area off the coast of Rosignano is also suggested and supported by satellite-derived SST observations \footnote{Estimates provided by JPL MUR MEaSUREs Project 2015 GHRSST Level 4 MUR Global Foundation Sea Surface Temperature Analysis version 4.1, see \url{https://podaac.jpl.nasa.gov/dataset/MUR-JPL-L4-GLOB-v4.1}, link visited in September 2025.}, which are higher than the ECMWF data. The analysis of such data highlight a positive SST anomaly, greater than 1 K, over a large part of the Northern Tyrrhenian and Ligurian Sea with respect to the 2003-2014 climatological average value (data not shown). The second sensitivity test (hereafter referred to as \texttt{SOLVA}) regarding SST is suggested by the data provided by Solvay Chimica Italia, the industrial chemical facility situated a few hundred meters inland from Rosignano and responsible for releasing warm water into the adjacent sea. These data include the temperature and discharge rate from the industrial facility. Figures \ref{fig:Solvay}a and \ref{fig:Solvay}b display the discharge rate and temperature, respectively, for the month of September 2020. The vertical black line indicates midnight of the 25th of  September 2020 (likely local time, though this point is not clear in the communication received from Solvay Chimica Italia). Looking closer to Figure \ref{fig:Solvay}, we can hypothesise that the abrupt shift of discharge rate, shortly after the black vertical line, is possibly linked to the closure of the Solvay channel because of adverse weather conditions in the Rosignano area. The average discharge channel temperature is approximately 28.7$^\circ$ C until midnight on September 25 (around 28.2$^\circ$ C for the entire month of September 2020). Therefore, during the \texttt{SOLVA} test, an initial SST value of 28.5$^\circ$ C was assumed, which corresponds, on average, to an increase of about 5$^\circ$ C compared to the SST provided by the global model. We acknowledge that it may be an overestimation but we claim it is helpful to provide an idea on how the emission of warm water into the sea may affect the development of the tornado-spawning supercell. To change the SST values in the inner grid, we assumed, subjectively, that the discharge channel influences the temperature of the sea water comprised in the white trapezoid shown in Figure \ref{fig:domain100}. To support such setting, we designed the trapezoid using few satellite images and looking at the portion of sea where the colour is significantly brighter than surrounding areas. In fact, it is known that the outflow of the industrial facility contains chemical substances that clarify the water. Additionally, both in the \texttt{SST15} and \texttt{SOLVA} simulations, the sea/atmosphere interface module was activated in the Meso-NH code, considering the sea as a single layer according to the formulation of \cite{gaspar1990simple}. In this way, the atmospheric model is coupled with a simple one-dimensional oceanographic model, and energy exchanges between the sea and atmosphere take into account the coupling. This is a fairly common and simple setup in numerical reconstructions of severe events involving ocean/atmosphere interactions, and it has been shown to yield valid results even in the Mediterranean Sea (see, for example, \cite{ricchi2021simulation}).
%
Finally, we chose the ECOCLIMAP v2.6 (resolution of approximately 1 km) as landuse dataset instead of the ECOCLIMAP Second Generation (resolution of approximately 300 m). In fact, although older, ECOCLIMAP v2.6 presents in the Rosignano area a pixel classified as ``Industries and commercial areas'', indicating the presence of an industrial site (presumably the Solvay Chimica Italia facility), whereas ECOCLIMAP Second Generation has generic urban classes in the same area. However, we acknowledge that no sensitivity tests were conducted to support such choice.

%
\subsection{Meteorological parameters}\label{sec:predictability}
The short duration (typically lasting only a few minutes) and small spatial scales (often tens to hundreds of meters) of tornadoes are below the grid spacing and effective resolution of current operational weather prediction models. Despite these limitations, several recent scientific studies \cite{matsangouras2014numerical,matsangouras2016study,hon2021observation,avolio2021multiple,avolio2022tornadoes,de2024conceptual,de2025significant} have employed numerical models to investigate the development of tornado-spawning supercells. These studies  primarily focused on identifying the weather conditions, both at the synoptic scale and mesoscale, that are conducive to the formation, growth, and intensification of these events. \cite{bagaglini2021synoptic} identified a set of atmospheric synoptic and mesoscale precursors as robust (i.e., statistically significant) factors in the generation of tornadoes in different Italian regions. In Appendix \ref{sec:appendix}, equations \ref{eq:ws}-\ref{eq:lcl}, we provide a concise summary of such relevant indices, namely: low level shear (LLS), storm relative helicity in the 0-1 km layer (SRH1km), an equivalent measure of the convective available potential energy (WMAX), and the lifting condensation level (LCL). Table \ref{tab:pre-meso} summarizes the statistical distributions of these atmospheric precursors during tornado events classified as EF1 and EF2+, as reported by \cite{bagaglini2021synoptic}. In our numerical experiments, the same indices were extracted at the grid point closest to the event location. While the higher spatial resolution of our data doesn't allow for a direct comparison with those in \cite{bagaglini2021synoptic}, these values still provide an indication of the presence of environments conducive to tornado development.

%
\section{Results}\label{sec:results}
To assess whether the model simulations adequately reconstruct the atmospheric environment conducive to tornadogenesis, we first present the spatial distribution of the mesoscale precursors identified by \cite{bagaglini2021synoptic}. In Appendix \ref{app:fig2}, Figures \ref{fig:ex2011-12-17}-\ref{fig:ex2020-09-25}, we illustrate maps of the LLS, SRH1km, WMAX and LCL from the Meso-NH simulations at the presumed time of waterspout occurrence; such maps are drawn using data from the inner domain with a grid spacing of 500 m. Figures \ref{fig:ex2011-12-17}-\ref{fig:ex2020-09-25} serve as ``snapshots'' of the atmospheric conditions at the estimated time of the event. The visual inspection of the model outputs, suggests that the indices are compatible with severe phenomena. 
This statement is further supported by the maps of the updraft helicity (UH) in the 2-5 km layer for the four cases (plots not shown). UH is a diagnostic variable designed to quantify updraft rotation in simulated storms \cite{kain2008some,clark2013tornado}, and hourly maximum UH values can serve as a useful proxy for tornado occurrence when considered in conjunction with environmental conditions conducive to tornadogenesis \cite{gallo2016forecasting}. In our simulations performed with a grid spacing of 500 m, UH values exceeding 50 ${m^2}/{s^2}$, a commonly adopted threshold for tornado forecasting \cite{clark2012forecasting}, are found in all four cases.
As regards mesoscale parameters, for the 17DEC2011  case (Figures \ref{fig:ex2011-12-17}a-d), we can observe high values of LLS (approximately 6-8 ${m}/{s}$). WMAX values are relatively low (less than $36$ ${m}/{s}$), but this is consistent with the period (mid December) when it is known \cite{lolis2024variability} that the energy available for convection reaches its annual  minimum. Conversely, elevated SRH1km values (exceeding 100 ${m^2}/{s^2}$) are simulated within 20 km of the point of interest. For the 27NOV2012   case, the Rosignano Solvay area is characterised by even higher values of SRH1km (greater than $200$ ${m^2}/{s^2}$; Figure \ref{fig:ex2012-11-27}b) and high values of WMAX (greater than $70$ ${m}/{s}$) in a sea portion close to the point of interest (distance less than 12 km, see Figure \ref{fig:ex2012-11-27}b). As expected, considering the limited resolution of the numerical  models, the values of the wind gusts at 10 m are not indicative of particularly extreme events (maps not shown); indeed, the wind gust is less than 90 $km/h$ (the threshold for an IF0 event according to the International Fujita scale). As discussed earlier, the 10SEP2017    case (Figures \ref{fig:ex2017-09-10}) affected a large part of the Tyrrhenian coasts with widespread severe weather conditions both north and south of the area of interest \cite{federico2019impact}. The values reconstructed by the model are consistent with the conditions conducive to severe weather events: LCL values (Figure \ref{fig:ex2017-09-10}d) in the range 400-600 $m$ above ground level, WMAX values greater than 60 $m/s$ (see Figure \ref{fig:ex2017-09-10}b), and wind gusts, probably associated with thunderstorm downdrafts, with intensities exceeding 120 $km/h$ (plot not shown). However, for this case study the initial conditions (i.e., analysis) provided by the global model lead to high-resolution simulations that produce errors both in the position and intensity of precipitations \cite{capecchi2021assimilating,federico2019impact}. It is only through the assimilation of radar observations that it is possible to reproduce correctly the low-level flow and the regions of convergence, responsible for the initiation of thunderstorms and the resulting squall winds \cite{federico2019impact}. Finally, for the 25SEP2020     case, for which the ESWD database reports two cases of ``tornado/waterspout'' with intensity IF2 and IF1 in the Rosignano area, maps are similar to those produced for the 10SEP2017    case (both cases occur in September), but with lower values for WMAX (in the range 52-56 ${m}/{s}$, see Figure \ref{fig:ex2020-09-25}b), since this event is characterised by a north-westerly flow, rather than southerly, thus cooler and drier air is advected in the low-levels. Further, buoyancy in the 0-1 km layer is not extreme, SRH1km values are pretty low approximately 25 ${m^2}/{s^2}$, see Figure \ref{fig:ex2020-09-25}d, and the LCL is relatively high (cloud base height approximately 1400 m, see Figure \ref{fig:ex2020-09-25}c). Table \ref{tab:atmo-prec} presents a comprehensive summary of the atmospheric precursors from the numerical reconstructions, reporting the average values of the model outputs from the four grid points nearest to the point of interest.
\begin{table}
\caption{Average values of selected atmospheric variables from numerical reconstructions, computed over the four grid points nearest to the point of interest for each case study.}\label{tab:atmo-prec}
  \centering
  \begin{tabular}{lllll}
    \toprule
\textbf{Variable [unit]} & \textbf{17DEC2011} & \textbf{27NOV2012} & \textbf{10SEP2017} & \textbf{25SEP2020} \\
    \midrule
LLS [$m/s$]               & 7.9       & 8.7       & 9.0       & 6.0     \\
SRH1km [$\frac{m^2}{s^2}$]             & 130.8     & 248.3     & 351.0     & 26.9     \\
WMAX [$m/s$]              & 35.9      & 71.5      & 62.9      & 55.2     \\
LCL [$m$]               & 1180      & 468       & 480       & 1428     \\
    \bottomrule
  \end{tabular}
\end{table}
%

\begin{figure}[t]
\includegraphics[width=14cm]{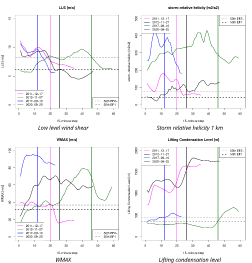}
  \caption{For each event (colours indicated in the legend on the left), we show the temporal evolution of mesoscale precursors listed in Table \ref{tab:pre-meso}. Data are extracted from the model simulation with a grid spacing equal to 500 m. The horizontal dashed lines indicate the median of mesoscale precursors for EF1 and EF2+ events, as in \cite{bagaglini2021synoptic}. The continuous vertical lines indicate the presumed time of the event. On the x-axis, the temporal increment is shown in 15-minute units from the model initialisation time.}\label{fig:ProfilePrecursori}
\end{figure}

%
To provide a more comprehensive understanding of the temporal evolution of the model outputs, Figure \ref{fig:ProfilePrecursori} presents time series of the mesoscale precursors listed in Table \ref{tab:pre-meso}, extracted within a region encompassing Rosignano Solvay, which corresponds to the domain depicted in Figure \ref{fig:domain100} with the red square. The dashed (dotted) black horizontal lines denote the 50th percentile for EF1 (EF2+) cases identified by \cite{bagaglini2021synoptic}. Solid vertical lines indicate the reported time of the event, sourced from data gathered from the ESWD database and listed  in Table \ref{tab:startdates}. Upon careful examination of the plots in Figure \ref{fig:ProfilePrecursori}, we observe that for certain events (e.g., the 27NOV2012   event, highlighted in dark green), the maximum precursor values are reached some time prior to the presumed event time, approximately two hours earlier. For the 27NOV2012   and 10SEP2017    events, the parameter values are particularly high, comparable to those typically associated with EF2+ events. Specifically, the LLS reaches approximately 8.7 $m/s$ for 27NOV2012   and 9.0 $m/s$ for 10SEP2017. Additionally, the SRH1km is approximately 298 and 351 $m^2/s^2$, respectively, while WMAX reaches 71.5 and 92.9 $m/s$. The LCL is less than 500 m in both cases; this similarity is unsurprising, as the two cases share common synoptic and mesoscale features (an intense meridional and humid flow), as described in Section \ref{sec:data}.

\begin{figure}[t]
\includegraphics[width=14cm]{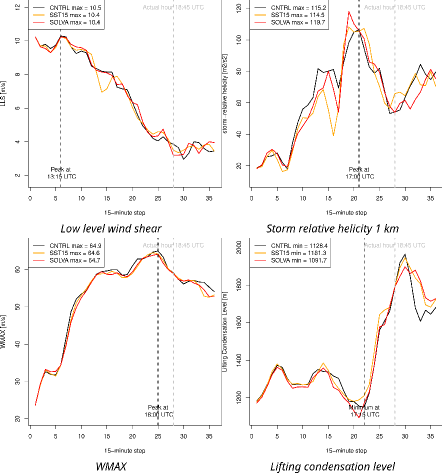}
\caption{25SEP2020     case: mesoscale precursors extracted from the \texttt{CNTRL} (black line), \texttt{SST15} (orange line), and \texttt{SOLVA} (red line) simulations. The dashed grey vertical line indicates the actual time of the event reported by the ESWD database, whereas the dashed black vertical line indicates the times with the extremes values of the environmental parameters as reconstructed by the CNTRL simulation (maximum value for LLS, SRH1km and WMAX, minimum for LCL).}\label{fig:testSST}
\end{figure}

For the 25 September 2020 case, the results of the two sea surface temperature sensitivity experiments (\texttt{SST15} and \texttt{SOLVA}) were compared against the control simulation (\texttt{CNTRL}), in which SST fields were prescribed from the driving global model. In the \texttt{SST15} (\texttt{SOLVA}) simulation, the initial SST field was increased by approximately 1.5 $K$ (5 $K$) over the sea portion shown in Figure \ref{fig:domain100} with the white trapezoid. To compare the sensitivity experiments with the control simulation, Figure \ref{fig:testSST} presents the temporal evolution of the mesoscale precursors. In all three simulations, the model successfully reconstructs the event but predicts the timing of the peak too early relative to the observed occurrence. In other words, the instability parameters reach their peak (the minimum value in the case of LCL) earlier in the simulation than in the observations.
The analysis of Figure \ref{fig:testSST} indicates that the modified SSTs do not influence significantly the outputs of both \texttt{SST15} and \texttt{SOLVA} simulations with respect to the \verb+CNTRL+ run. 
In fact, when we extract the four grid points closest to the point of interest at the presumed time of the event and compute the median value, we find that the \texttt{SOLVA} simulation provides higher values for all variables except WMAX; however, these increments are modest (less than 3\%; data not shown).

%
\section{Discussion}\label{sec:discussion}
Using the limited-area atmospheric model Meso-NH, we reconstructed the mesoscale meteorological conditions associated with four severe events that occurred over the past 15 years in the Rosignano Solvay area along the Tyrrhenian coast of Italy. For the 25SEP2020     event, two SST sensitivity tests were performed, where, in a limited sea area in front of Rosignano Solvay (Figure \ref{fig:domain100}), the temperature was increased by 1.5 K and 5 K, respectively. For two out of the four cases under examination, namely 27NOV2012   and 25SEP2020, the ESWD database report the occurrence of a ``tornado/waterspout'' in the Rosignano Marittimo area. The first case is assigned category IF1, and the second case category IF2, according to the International Fujita (IF) Scale classification scheme. The 10SEP2017    case, does not report data on the Rosignano Solvay area but rather the ``intense rainfall'' phenomenon less than 10 kilometres to the north \cite{capecchi2021assimilating}. For the 17DEC2011  case, no data are available in the ESWD database. However, meteorological reports archived by the LaMMA Laboratory, along with the regional criticality bulletin issued the previous day, indicate widespread adverse weather conditions across the region. These included strong winds over the Tuscan Archipelago and a severe storm surge north of Elba Island. Notably, the nearest buoy, located near Gorgona Island approximately 45 km from Rosignano Solvay, recorded a significant wave height exceeding 6.5 meters. These considerations suggest that for this case the damages in the Rosignano area are likely linked to strong wind gusts associated with thunderstorm lines rather than the occurrence of a waterspout making landfall.

Owing to the restricted capability of limited-area numerical weather prediction models to accurately forecast tornadoes and waterspouts (e.g.,  \cite{bagaglini2021synoptic} and \cite{ingrosso2020statistical}), we have considered the meteorological variables that have been shown to be robust precursors for the origin, development, and intensification of tornadoes/waterspouts \cite{bagaglini2021synoptic}. The analysis of the model outputs can be conducted by extracting mesoscale precursor values from an area near the point of interest and examining their temporal evolution. A comparison between the peak values of instability parameters and the observed event timing reveals that the model tends to predict the peak by approximately 2 to 3 hours too early. This is evident, for example, considering the variable WMAX in Figure  \ref{fig:ProfilePrecursori}c for the 27NOV2012   (red line) and 25SEP2020     (black line).

To evaluate the possible role played by SSTs in the waterspout occurrences, we focused our analysis on two sensitivity numerical experiments. It is known \cite{miglietta2017effect,avolio2021multiple,avolio2022tornadoes} that higher SSTs imply a warmer and moister atmosphere in the lower layers, therefore favourable to more unstable conditions and potentially stronger upward motions. Previous studies  \cite{miglietta2017effect} have shown that relatively small variations in SST values ($\pm$ 1 $K$) can significantly suppress or enhance the vigour of a supercell capable of giving rise to tornado phenomena. For the sole case of 25SEP2020, in addition to the \texttt{CNTRL} simulation, where SSTs are deduced from the global model, we performed two test simulations to verify the influence of warmer SSTs on the model results. In the first test (\texttt{SST15}), the SST of the sea area in front of the fraction of Rosignano Solvay was increased, on average, by +1.5 $K$ compared to the value provided by the global model. In the \texttt{SOLVA} test, the SST increment was set to +5 $K$. This value, seemingly high compared to tests present in the literature, is suggested by the data provided by Solvay Chimica Italia, which report an average temperature of the discharge canal waters for the month of September 2020 (up to and including September 24) of approximately 28.7$^\circ C$. In Figure \ref{fig:testSST}, we show the values of mesoscale precursors extracted in a box containing the point of interest for the three simulations \texttt{CNTRL}, \texttt{SST15}, and \texttt{SOLVA}. The data were extracted considering the inner grid shown in Figure \ref{fig:domain100}, which has a grid spacing of 100 m. If we extract the data at the observed time of the event $\pm 2$ hours,
we can conclude that the two test simulations do not significantly deviate from \texttt{CNTRL}, as the median values and 95th percentiles of SRH1km in the \texttt{SOLVA} simulation are only slightly higher than those provided by \texttt{CNTRL}. In fact, the sea area where the SSTs were modified in the \texttt{SST15} and \texttt{SOLVA} simulations is too small to exert a significant influence on the atmospheric variables even locally.

\begin{figure}[t]
     \includegraphics[width=14cm]{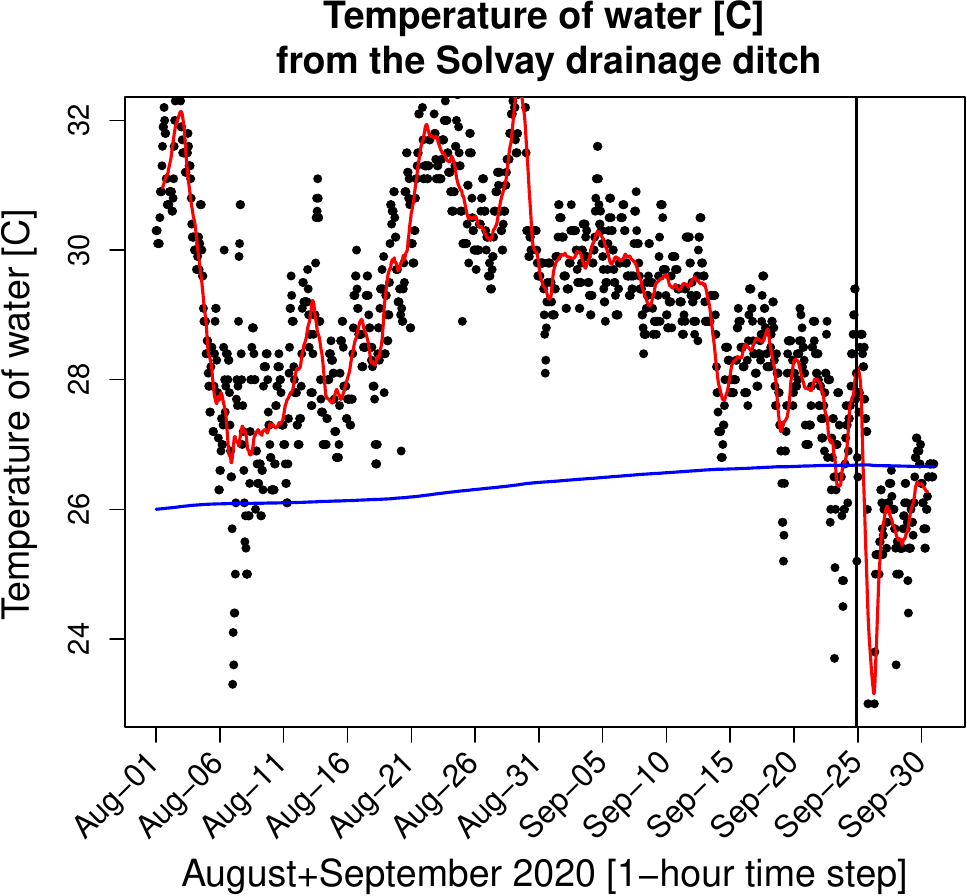}
     \caption{August/September 2020: according to equation (\ref{eq:appro1}), the blue line indicates the variation of sea surface temperature within the trapezoid in front of the fraction of Rosignano Solvay due to the effect of the Solvay plant discharge water, whose values are shown with the black points (hourly data) and red line (24-hour running mean).}\label{fig:eqappro}
\end{figure}

\begin{figure}[t]
     \includegraphics[width=14cm]{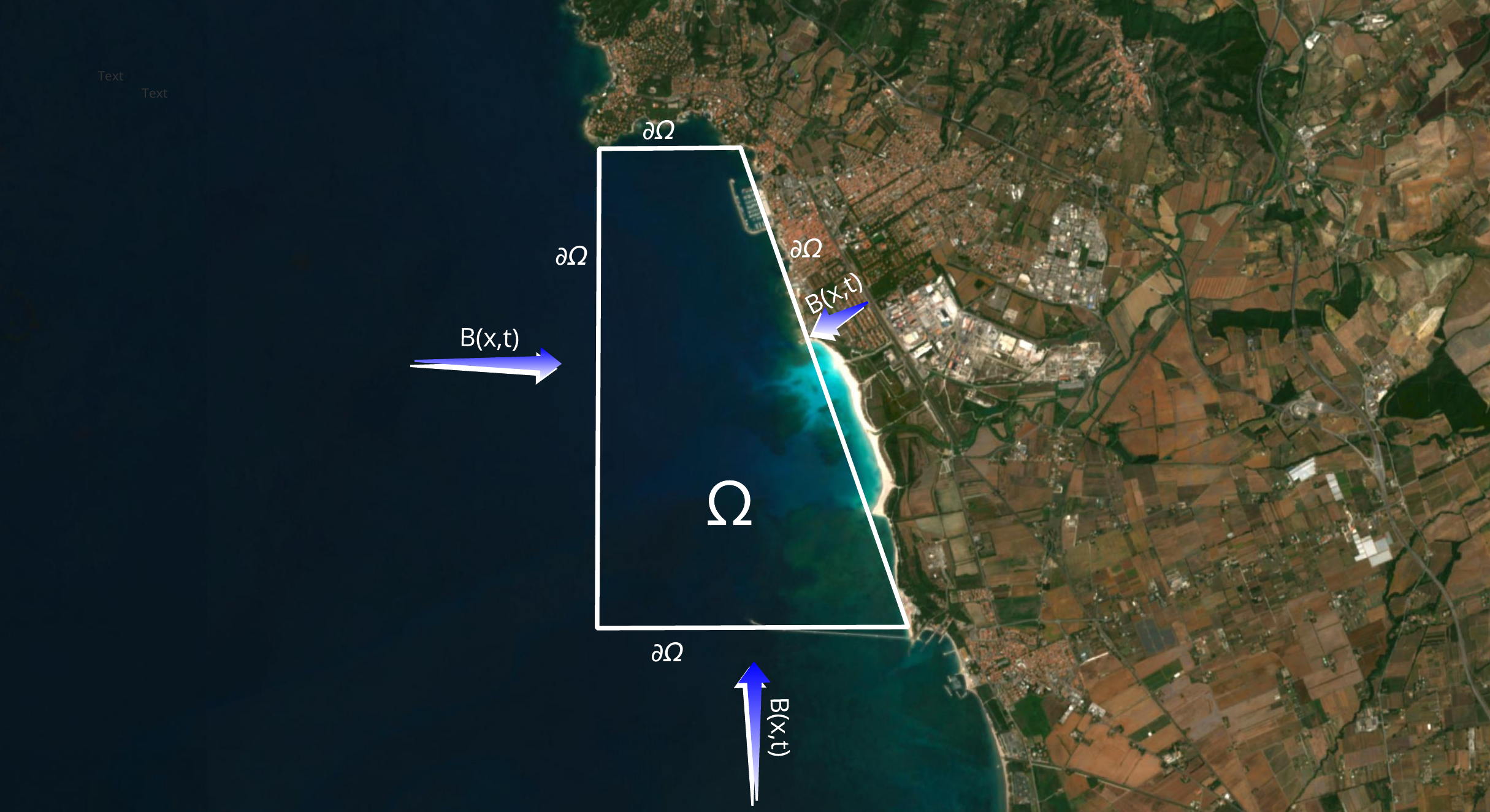}
     \caption{Sketch of the horizontal fluxes $\mathbf{B}(\mathbf{x},t)$ affecting the boundaries of the sea portion facing the Rosignano Solvay town and indicated with the white trapezoid $\Omega$. Satellite image from Sentinel-2 cloudless \textcopyright\ EOX IT Services GmbH.}\label{fig:idealmodel}
\end{figure}

It is evident that there remains a high degree of uncertainty regarding the portion of the sea whose surface temperature is influenced by the temperature of the discharge water. With the choice of the trapezoid shown in Figure \ref{fig:domain100}, we assumed that the volume of the water affected by the discharge channel is approximately 50 million m$^3$. Indeed, the area of the trapezoid is about 10 km$^2$, and the average depth is approximately 5 m (this is the bathymetry of the assumed area from the ETOPO database, \cite{etopo}). As a rough working hypothesis, we assume that there are no exchanges between the water inside the trapezoid, denoted by $\Omega$ in Figure \ref{fig:idealmodel}, and the surrounding waters (i.e., we assume that $\Omega$ is a closed system with respect to the open sea). We can then assume that its surface temperature $T(t)$, varying over time $t$ (for simplicity, we consider discrete time intervals, $t\in\mathbb{N}$), follows the relationship:
\begin{equation}
T(t)=T(t-1)+\alpha\frac{R(t)}{m_0}\Delta T_S(t-1), \label{eq:appro1}
\end{equation}
where $R(t)$ is the flow rate at time $t$ through the boundary of $\Omega$ (the flow rate for the month of September 2020 is shown in Figure \ref{fig:Solvay}a), $m_0$ is the total mass of water inside the trapezoid $\Omega$ (approximated to 50 million tons), $\Delta T_S(t-1)$ is the difference between the temperature of the discharge channel water $T_S$ and the temperature of the trapezoid water $T$ at time $(t-1)$, and $\alpha$ is a constant with dimensions of time. Assuming for simplicity the initial time as 00:00 local time on August 1, 2020, and setting $T(0)=26.0^\circ  $ $C$ (an approximate value close to that measured by the closest buoy), we find that the sea surface temperature within the trapezoid increases by approximately $0.7^\circ$ $C$ in almost two months to reach 26.68$^\circ$ $C$. In Figure \ref{fig:eqappro}, the blue line shows the evolution of $T(t)$ as computed using equation (\ref{eq:appro1}), with an initial temperature of $T(0)=26.0^\circ$ $C$ on August 1, 2020.
However, it is evident that the sea surface temperature contained within the trapezoid shown in Figure \ref{fig:domain100} cannot be influenced solely by the discharge water but also by the inflow/outflow with the surrounding sea. In a more general model, we should also consider the vertical flows (evaporation/precipitation) through the surface due to the sea/atmosphere interface, which we considered negligible. Modifying the original hypothesis, if we also consider the temperature variation due to an inflow (equal to that of the discharge channel) and the sea surface temperature of the surrounding waters, equation (\ref{eq:appro1}) could be modified to:
\begin{equation}
T(t)=T(t-1)+\alpha\left(\frac{R(t)}{m_0}\Delta T_S(t-1) + \frac{R(t)}{m_0}\Delta T_M(t-1)\right) \label{eq:appro2}
\end{equation}
where $\Delta T_M(t-1)$ is the difference between the temperature of the surrounding sea $T_M$ and that of the trapezoid. More generally, the variation of sea surface temperature within the trapezoid is given by:
\begin{equation}
T(t)=T(t-1)+\alpha\int_{\partial \Omega}\mathbf{B}(\mathbf{x},t)\cdot \mathbf{n}\ ds \label{eq:appro3}
\end{equation}
where $\partial \Omega$ is the boundary of $\Omega$ (the edges), \textbf{B} is the vector field that takes into account the inflow and outflow of the sea water and the temperature difference through $\partial \Omega$ at time $t$, and $\mathbf{n}$ is the  outward-pointing normal vector to $\partial \Omega$. A schematic representation of the fluxes through the trapezoidal domain is shown in Figure \ref{fig:idealmodel}.
In the practical case at hand, in equation (\ref{eq:appro1}), we assumed that $\mathbf{B}(\mathbf{x},t)$ is equal to $\frac{R(t)}{m_0}\Delta T_S(t-1)$ and that the boundary of $\Omega$ consists of a single point. In the general case, estimating \textbf{B} is difficult, so equation (\ref{eq:appro3}) is not applicable in general, despite its greater rigour.

Conducting a model simulation of marine currents and investigating variations in the horizontal and vertical thermal gradient of the waters in the sea area in front of the fraction of Rosignano Solvay is beyond the scope of this study. However, based on the scientific evidence present in the literature \cite{miglietta2017effect,avolio2021multiple,avolio2022tornadoes}, it is reasonable to expect a more pronounced influence of the SST anomaly on atmospheric instability indices only if we were to consider a much larger area than the trapezoid $\Omega$ shown in Figure \ref{fig:idealmodel}, which has an area of approximately 10 $km^2$. This conclusion further highlights the importance of an accurate and region-specific characterization of sea surface temperature in order to properly assess its impact on atmospheric processes. Such information could be obtained through the integration of high-resolution satellite observations (e.g., infrared and microwave SST products) and in situ measurements (e.g., buoys or coastal monitoring stations), and subsequently assimilated or prescribed as boundary conditions in coupled atmosphere–ocean or high-resolution atmospheric numerical simulations.

\section{Conclusions}\label{sec:conclusions}
As a general consideration, we can state that the initiation and evolution of waterspouts in the Tyrrhenian coast of Tuscany seem to be linked to large-scale meteorological configurations rather similar to each other. This element tends to rule out causes related to local situations. Further, the \texttt{SST15} and \texttt{SOLVA} sensitivity experiments demonstrate that modifying the SSTs in a small portion of sea has a marginal impact on instability parameters. However, the fact that the affected area is always the same may suggest that the repetition of the occurrences are not entirely random. One possible explanation could be related to the morphology of the coastline and of its immediate surroundings characterised by different landuses (both dense constructions and cultivated fields). To corroborate the possible influence of the orographic effect on the development of waterspouts in the Rosignano Solvay area, numerical experiments will be performed by reducing or removing the model topography as done previously by \cite{matsangouras2014numerical} and \cite{miglietta2017numerical}. 

Further studies are required to assess whether the shape of the coastline, the morphology of the adjacent inland areas, and the presence of nearby islands, such as Corsica and the Tuscan Archipelago to the west, play a role in the onset and development of waterspouts in the Rosignano Solvay area. Another possible line of research concerns the investigation of the role of urban and anthropogenic heat release in the onset and development of waterspouts in the Rosignano Solvay area. Sensitivity simulations incorporating urban parameterizations and anthropogenic or industrial heat fluxes could help assess potential modifications to the near-surface energy balance and boundary-layer structure, as highlighted by \cite{sari2023understanding}. In particular, urban heat release enhances sensible heat fluxes, increases low-level buoyancy, and locally intensifies vertical motions, thereby potentially contributing to convective development. In contrast, the contribution of the heated water discharged into the limited sea area via the outflow channel appears relatively minor (Figure \ref{fig:eqappro}), and its impact on the elevated frequency of waterspouts in this region cannot be reliably evaluated based on the numerical experiments conducted here.

\vspace{6pt} 




\subsection*{Acknowledgments}
The authors would like to thank the Municipality of Rosignano Marittimo for stimulating interest in this study and for providing observations and photographic documentation that supported the analysis of the weather events.

The authors are also grateful to Solvay Chimica Italia for supplying flow rate and temperature data (Figures \ref{fig:Solvay}a-\ref{fig:Solvay}b) from the discharge ditch, commonly known as ``fosso bianco'', for the months of August and September 2020.

Finally, the authors acknowledge the ECMWF reanalysis for the ERA5 data, available from the Copernicus Climate Data Store \url{https://cds.climate.copernicus.eu}.
\clearpage

\section[\appendixname~\thesection]{Appendix A}\label{app:fig1}
In the figures below, we show, for each case, the ERA5 data for some synoptic parameters, namely: 250-hPa wind speed, mean sea level pressure, and deep level shear.
\begin{figure}[h]
\includegraphics[width=14cm]{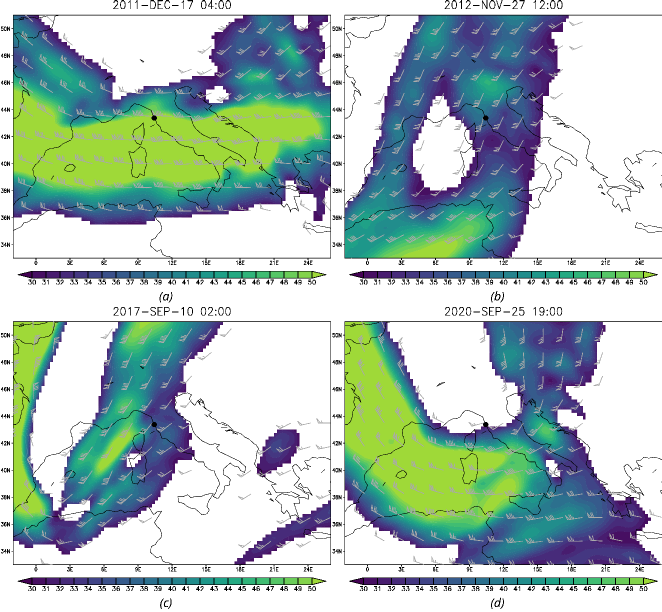}
\caption{ERA5 data 250-hPa speed (unit in knots) and direction at the approximate time of the waterspout for the 17DEC2011  (a), 27NOV2012   (b), 10SEP2017    (c) and 25SEP2020     (d) case. Barbs are shown for wind speed greater than 30 knots. The approximate location of Rosignano Solvay is indicated with the black point ($\bullet$).}\label{fig:250jetERA5}
\end{figure}
%
\begin{figure}[h]
\includegraphics[width=14cm]{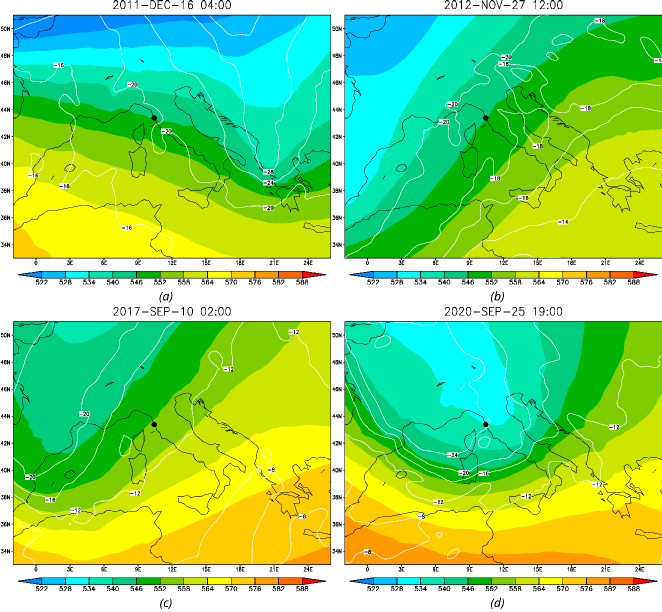}
\caption{As in Figure \ref{fig:250jetERA5} but for geopotential height at 500-hPa pressure level (unit in dam) with superimposed temperature (unit in degree Celsius, contour).}\label{fig:hgt500ERA5}
\end{figure}
%
\begin{figure}[h]
\includegraphics[width=14cm]{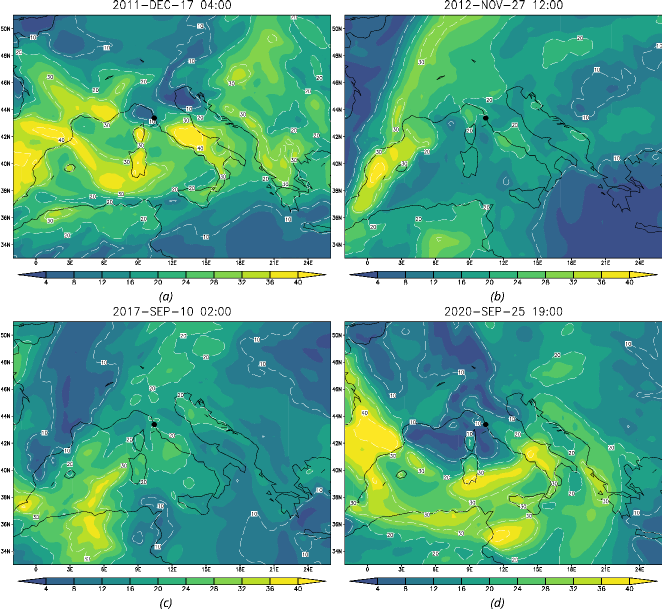}
\caption{As in Figure \ref{fig:250jetERA5} but for deep layer shear in the 0-6 km layer (unit in $m/s$).}\label{fig:dlsERA5}
\end{figure}
%
%
\clearpage

\section[\appendixname~\thesection]{Appendix B}\label{sec:appendix}
\renewcommand{\thetable}{\thesection\arabic{table}}
\setcounter{table}{0}
\renewcommand{\theequation}{\thesection\arabic{equation}}
\setcounter{equation}{0}
List of the four mesoscale instability precursors that \cite{bagaglini2021synoptic} identified as significant in the generation of tornadoes in different Italian regions:
\begin{itemize}
\item[-] Vertical wind shear at three different levels, namely Deep Layer Shear (DLS; in the 0-6 km layer), Mid-Level Shear (MLS; in the 0-3 km layer), and Low-Level Shear (LLS; in the 0-1 km layer). The vertical wind shear at pressure level $p$ is defined as:
\begin{equation}\label{eq:ws}
WS_p =\mid\mid\mathbf{u}_p - \mathbf{u}_{10m}\mid\mid
\end{equation}
where $\mathbf{u}_p$ and $\mathbf{u}_{10m}$ are the horizontal wind vectors at pressure level $p$ and at 10 m above ground level, respectively, and $\mid\mid\cdot\mid\mid$ indicates the Euclidean norm. Often, $p$ takes values of 900-, 700-, and 500-$hPa$ for LLS, MLS, and DLS, respectively. The unit of measurement is $m/s$;
\item[-] Storm-relative helicity in the 0-1 km layer (SRH1km). Defined according to \cite{evans2001examination} as:
\begin{equation}\label{eq:srh1km}
\textrm{SRH1km} = \int_{0\ km}^{1\ km} (\mathbf{v} - \mathbf{c}) \cdot \textrm{d}\omega,
\end{equation}
where $\mathbf{v}=(u,v)$ is the horizontal wind vector, $\mathbf{c}$ is the storm motion vector, and $\omega$ is the horizontal vorticity vector associated with vertical wind shear, given by $(\partial w/\partial y - \partial v/\partial z,\partial u/\partial z - \partial w/\partial x)$. The unit of measurement is $m^2/s^2$;
\item[-] Convective Available Potential Energy ($CAPE$), defined as:
\begin{equation}\label{eq:cape}
CAPE = \int_{LFC}^{EL} B \, \textrm{d}z
\end{equation}
where $z$ is altitude, $B$ is buoyancy, $LFC$ is the level of free convection, and $EL$ is the equilibrium level. For the reasons indicated in \cite{bagaglini2021synoptic}, $WMAX = \sqrt{2 \cdot CAPE}$ is more frequently used than $CAPE$. The unit of measurement of $WMAX$ is $m/s$;
\item[-] Lifting Condensation Level ($LCL$), indicates the height (in $m$) at which an air parcel forced to ascend from the surface reaches saturation level (relative humidity equal to 100\%). It can be considered as a good approximation of the cloud base height. It can be calculated using the approximate formula reported in \cite{lawrence2005relationship}, namely:
\begin{equation}\label{eq:lcl}
LCL \approx 125(T_{10m} - Td_{10m})
\end{equation}
where $T_{10m}$ and $Td_{10m}$ are the temperature and dewpoint temperature, respectively, at 10 m above ground.
\end{itemize}
%
\begin{table}[h]
\caption{Statistical distribution (5th, median, and 95th percentiles) of mesoscale precursors in the case of (a) EF1 or (b) EF2+ tornadoes/waterspouts. Data are sourced from \cite{bagaglini2021synoptic}.}\label{tab:pre-meso}  \centering
  \begin{tabular}{llll}
    \toprule

\textbf{Variable} & \textbf{5th percentile} & \textbf{Median} & \textbf{95th percentile} \\
    \midrule
(a) EF1            &              &                  &               \\
LLS                & 1.1          & 6.1              & 13.9          \\
SRH1km             & -0.5         & 43.5             & 187.5         \\
WMAX               & 1.8          & 31.7             & 67.1          \\
LCL                & 109          & 358              & 1139          \\
(b) EF2+            &              &                  &              \\
LLS                & 2.5          & 8.2              & 13.1          \\
SRH1km             & 12.2         & 76               & 214.5         \\
WMAX               & 15.1         & 36.9             & 61.3          \\
LCL                & 174          & 438              & 1559         \\
    \bottomrule
  \end{tabular}
  \label{tab:table}
\end{table}
%
\section[\appendixname~\thesection]{Appendix C}\label{app:fig2}
\renewcommand{\thefigure}{\thesection\arabic{figure}}
\setcounter{figure}{0}
\setcounter{table}{0}
In the figures below, we show, for each case, the Meso-NH model output (grid spacing is 500 m) of the four mesoscale instability precursors identified by \cite{bagaglini2021synoptic} and listed previously in \ref{sec:appendix}. Each map refers to the approximate time of waterspout occurrence, according to the ESWD record.
\begin{figure}[h]
\includegraphics[width=14cm]{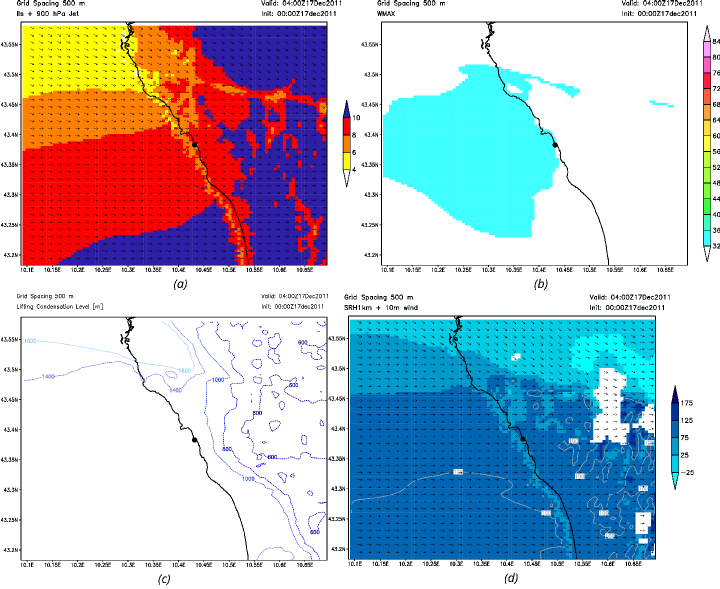}
\caption{outputs of th Meso-NH simulation at the presumed time of the waterspout event for the 17DEC2011  case: (a) low level shear, (b) WMAX, (c) lifting condensation level and (d) storm relative helicity in the 0-1 km layer. The approximate location of Rosignano Solvay is indicated with the black point ($\bullet$).}\label{fig:ex2011-12-17}
\end{figure}
%
\begin{figure}[h]
\includegraphics[width=14cm]{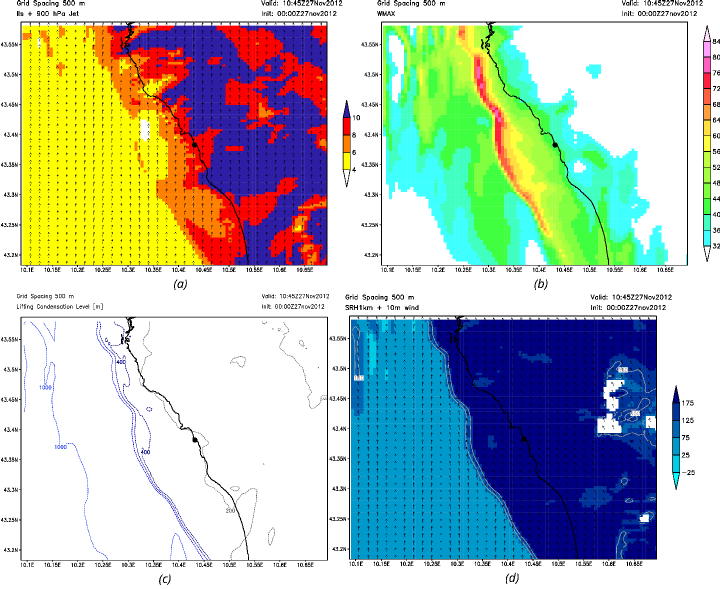}
\caption{As in Figure \ref{fig:ex2011-12-17} but for the 27NOV2012   case.}\label{fig:ex2012-11-27}
\end{figure}
%
\begin{figure}[h]
\includegraphics[width=14cm]{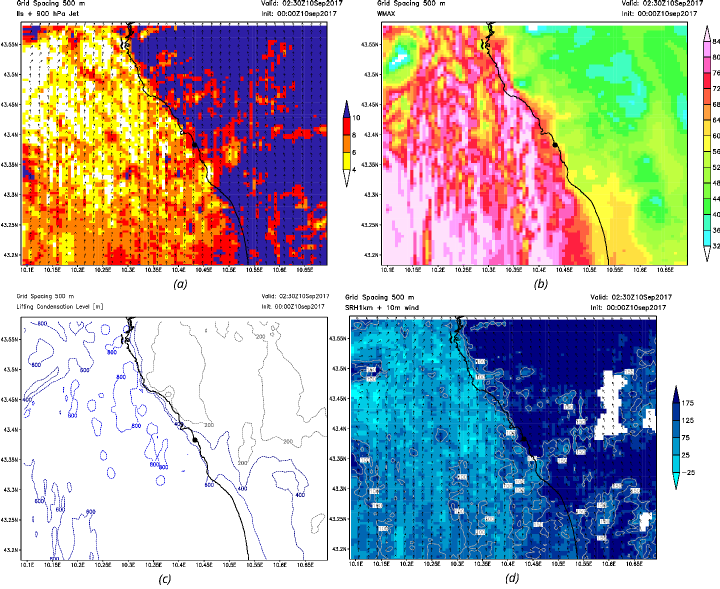}
\caption{As in Figure \ref{fig:ex2011-12-17} but for the 10SEP2017    case.}\label{fig:ex2017-09-10}
\end{figure}
%
\begin{figure}[h]
\includegraphics[width=14cm]{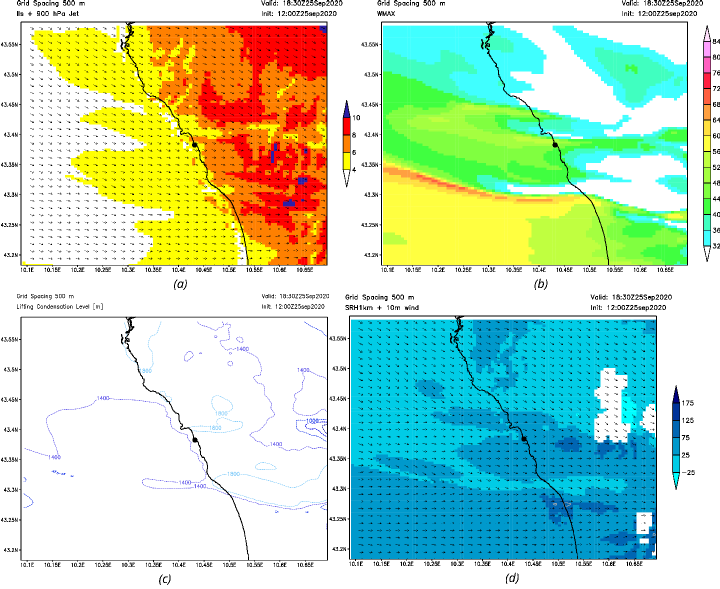}
\caption{As in Figure \ref{fig:ex2011-12-17} but for the 25SEP2020     case.}\label{fig:ex2020-09-25}\end{figure}
%


\clearpage
\bibliographystyle{unsrt}  
\bibliography{biblio_valcap}


\end{document}